\documentclass[iop]{emulateapj}
\usepackage{epstopdf}

\newcommand{\Msun}{{\rm M_{\odot}}}

\newcommand{\kmps}{\, {\rm km \, s^{-1}}}

\slugcomment{}

\shorttitle{CO($J=1-0$) Imaging of M51 with CARMA and NRO45}
\shortauthors{Koda et al.}

\begin{document}

\title{CO($J=1-0$) Imaging of M51 with CARMA and Nobeyama 45m Telescope}

\author{Jin Koda\altaffilmark{1,2},
Tsuyoshi Sawada\altaffilmark{3,4},
Melvyn C. H. Wright\altaffilmark{5},
Peter Teuben\altaffilmark{6},
Stuartt A. Corder\altaffilmark{2,7},
Jenny Patience\altaffilmark{2,8},
Nick Scoville\altaffilmark{2},
Jennifer Donovan Meyer\altaffilmark{1},
Fumi Egusa\altaffilmark{2}}

\email{jin.koda@stonybrook.edu}

\altaffiltext{1}{Department of Physics and Astronomy, Stony Brook University, Stony Brook, NY 11794-3800}
\altaffiltext{2}{Department of Astronomy, California Institute of Technology, Pasadena, CA 91125}
\altaffiltext{3}{Joint ALMA Office, Alonso de C´'{o}rdova 3107, Vitacura, Santiago 763-0355, Chile}
\altaffiltext{4}{National Astronomical Observatory of Japan, Mitaka, Tokyo, 181-8588, Japan}
\altaffiltext{5}{Department of Astronomy and Radio Astronomy Laboratory, University of California, Berkeley, CA 97420}
\altaffiltext{6}{Department of Astronomy, University of Maryland, College Park, MD 20742}
\altaffiltext{7}{North American ALMA Science Center, 520 Edgemont Road, Charlottesville, VA 22903}
\altaffiltext{8}{School of Physics, University of Exeter, Stocker Road, Exeter, EX4 4QL, Devon, UK}

\begin{abstract}
We report the CO($J=1-0$) observations of the  Whirlpool Galaxy M51 using both the Combined
Array for Research in Millimeter Astronomy (CARMA) and the Nobeyama 45m telescope (NRO45).
We describe a procedure for the combination of interferometer and single-dish data.
In particular, we discuss
(1) the joint imaging and deconvolution of heterogeneous data, (2) the weighting scheme based
on the root-mean-square (RMS) noise in the maps, (3) the sensitivity and uv-coverage requirements,
and (4) the flux recovery of a combined map.
We generate visibilities from the single-dish map and calculate the noise of each visibility based on the RMS noise.
Our weighting scheme, though it is applied to discrete visibilities in this paper, should be applicable
to grids in $uv$-space, and this scheme may advance in future software development.
For a realistic amount of observing time, the sensitivities of the NRO45 and CARMA visibility data sets are best matched by using the single dish baselines only up to 4-6 $k\lambda$ (about 1/4-1/3 of the dish diameter).
The synthesized beam size is determined to conserve the flux between
synthesized beam and convolution beam. 
The superior $uv$-coverage provided by the combination of CARMA long baseline data with 15 antennas
and NRO45 short spacing data results in the high image fidelity, which is evidenced by
the excellent overlap between even the faint CO emission and dust lanes in an optical {\it Hubble Space Telescope}
image and $PAH$ emission in an {\it Spitzer} $8\mu m$ image.
The total molecular gas masses of NGC 5194 and 5195 ($d=8.2 \rm \, Mpc$) are $4.9\times 10^9 \Msun$ and $7.8 \times 10^7 \Msun$,
respectively, assuming the CO-to-H$_2$ conversion factor of $X_{\rm CO}= 1.8\times 10^{20} \rm \, cm^{-2} [K \cdot km/s]^{-1}$.
The presented images are an indication of
the millimeter-wave images that will become standard in the next decade with CARMA and NRO45,
and the Atacama Large Millimeter/Submillimeter Array (ALMA).
 \end{abstract}


\section{Introduction}

Interferometers have an intrinsic limitation, namely, the problem of missing
information. An interferometer records the target Fourier components of
the spatial emission distribution, but an interferometer with
a  small number of antennas ($N$) can collect only a limited
number, $N(N-1)/2$, of the Fourier components instantaneously.
In addition, the finite diameter of each antenna limits the minimum
separation between antennas, which, in turn, imposes a maximum size on an object
that the interferometer can detect.
The zero-spacing data (i.e. zero antenna separation data) 
carry the important information of the total flux, and this
information is always missing.
The incomplete Fourier coverage ($uv$-coverage) also degrades
the quality of image.
Deconvolution schemes have been developed to extrapolate the observed
$uv$ data to estimate the missing information, however, the performance is
poor for objects with high contrast, such as spiral arms and the
interarm regions of galaxies.

The small-$N$ problem is  particularly severe
in millimeter astronomy, though it is greatly reduced with the 15-element Combined Array for
Research in Millimeter Astronomy (CARMA). CARMA combines the
previously independent Owens Valley Millimeter Observatory (OVRO)
array ($N=6$) and Berkeley-Illinois-Maryland Association (BIMA)
array ($N=10$ -- reduced to 9 for CARMA). The number of antenna pairs,
or {\it baselines}, is increased to 105 from the previous values of 15 (OVRO) 
and 45 (BIMA), providing a substantial improvement in  $uv$ coverage.
In most observatories, a few array configurations are used to increase
the number of baselines. The $uv$ coverage from one CARMA configuration
is equivalent to that from seven configurations with a 6-element array.
CARMA ensures the unprecedented $uv$ coverage 
in millimeter interferometry compared to previous mm-wave arrays.

Single-dish telescopes complement the central $uv$ coverage
and provide short baselines, including the zero-spacing baseline.
The combination of interferometer and single-dish data is not
trivial, though several methods have been suggested.
Existing methods can be categorized into three types.
The first method produces visibilities from a single-dish map
\citep{vog84, tak03} and adds single-dish and interferometer
data in the $uv$ domain.
\citet{pet10} discussed a mathematica formalism.
One issue faced when utilizing this method, the difficulties of which are discussed in \citep{hel03},
has been the weighting of the two sets of data in combination. \citet{rod08} and \citet{kur09}
manually set the single-dish weight relative to the weight of the interferometer to improve
the shape of the synthesized beam.
In this paper, we suggest a new weighting scheme based solely on the quality of the single-dish data. In our method, the single-dish weight is independent of the interferometer data and is intrinsic to the single-dish observations. It naturally down-weights (up-weights) the single-dish data when its quality is poor (high).
In the appendix, we discuss the sensitivity matching which makes the combination most effective.

The second type of combination method co-adds two sets of data in the image domain
\citep{sta99}. This approach produces a joint dirty image and synthesized beam
\footnote{The synthesized beam is the instrumental point spread function for the aperture
synthesis array; also know as the "dirty beam".}
by adding the single-dish map and the interferometer dirty image,
and single-dish beam and interferometer synthesized beam, respectively.
The joint dirty image is then deconvolved with the joint dirty beam.
This technique was adopted for the BIMA-SONG survey \citep{hel03}.
\citet{cor88}, and recently \citet{sta02}, also discussed a non-linear
combination technique through joint deconvolution with the maximum
entropy method (MEM).

The third method was introduced by \citet{wei01} and operates in the
Fourier plane. The deconvolved interferometer map and single-dish map
are Fourier transformed and then the central $uv$-space
from interferometer data is replaced with single-dish data.

This paper describes the observations, data reduction, and combination of
CARMA and NRO45 data of M51.
Our procedure unifies the imaging techniques for interferometer mosaic data,
heterogeneous array data, and the combined data of single-dish and interferometer.
Earlier data reduction and results have been published  \citep{kod09}.
The method and results are the same,
but we have re-calibrated and reduced the entire data set using higher
accuracy calibration data.
In \S \ref{sec:carma} and \S \ref{sec:nro45}, we describe the CARMA and NRO45
observations and calibration. 
The deconvolution (such as CLEAN) is
detailed in \S \ref{sec:imaging} for three 
cases: (a) homogeneous array, single-pointing observations,
(b) heterogeneous array, single-pointing observations, and
(c) heterogeneous array, mosaic observations.
The weighting scheme in co-adding
the images from a heterogeneous array (with multiple primary beams) is
discussed in \S \ref{sec:weighting}.
The result from this subsection is also essential for the combination of interferometer
and single-dish data. The conversion of
a single-dish map to visibilities is explored in \S \ref{sec:comb},
and \S \ref{sec:fidelity} discusses the resultant map and image fidelity.
A summary of the requirements of single-dish observations
for the combination are explained in \S \ref{sec:req}.
Comments on other combination methods are given in \S \ref{sec:othermethods}.
The summary is in \S \ref{sec:summary}, and 
sensitivity matching between single-dish 
and interferometer observations is discussed in Appendix \ref{sec:senmatch}.

\section{CARMA}\label{sec:carma}

\subsection{Observations} \label{sec:obscarma}

High resolution observations of the Whirlpool galaxy M51 in the CO($J=1-0$) line
were performed with the Combined Array for Research in Millimeter
Astronomy (CARMA) during
the commissioning phase and in the early science phase during the CARMA construction
(2006-2007).
CARMA is a recently-developed interferometer, combining the six 10-meter antennas of the Owens
Valley Radio Observatory (OVRO) millimeter interferometer and the nine 6-meter
antennas of the Berkeley-Illinois-Maryland Association (BIMA) interferometer.
The increase to 105 baselines provides  superior
$uv$-coverage and produces high image fidelity. The C and D array configurations are used. The baseline length spans over
30-350 m (C array) and 11-150 m (D array).

The observations started with the heterodyne SIS receivers from OVRO
and BIMA. The typical system temperature of these original receivers
was $\sim 200$ K in double-side band. The receivers of the 15 antennas were being
upgraded one antenna at a time during the period of observations, but
the process was not completed before these observations finished.
The system temperature of the new replacement receivers is typically $\sim 100$ K at 115 GHz.

The first-generation CARMA digital correlators were used as a spectrometer. They had
three dual bands (i.e. lower and upper side bands) for all 105 baselines.
Each band had  five configurations of bandwidth -- 500, 62, 31, 8, and
2 MHz --
which have 15, 63, 63, 63, and 63 channels, respectively.
We switched the configuration of band 1, 2, 3 between (band 1, 2, 3) = (500, 500, 500) 
for gain calibration quasar observations and (band 1, 2, 3) = (62, 62, 62) for target integrations.
This "hybrid" configuration ensures both a sufficient detection of the gain calibrator
1153+495 with the total 3 GHz bandwidth (i.e. 3 bands $\times$ 2 side bands $\times$
500MHz bandwidth) and a sufficiently wide velocity coverage for the main galaxy NGC 5194.
The total bandwidth is 149.41 MHz after dropping edge 6 channels at each side,
which could be noisier than the central channels.
The companion galaxy NGC 5195 was not included in the velocity coverage,
although it was detected in the NRO45 map (\S \ref{sec:obsnro45}).

The hybrid mode observations require a special calibration for amplitude and
phase offsets between bands and between configurations. We observed a bright
quasar by changing the correlator configurations in time sequence:
1. (band 1, 2, 3) = (500, 500, 500), 2. (62, 62, 62), 3. (500, 62, 62), 4. (62, 500, 62), and
5. (62, 62, 500). Each configuration spends 5 min on integration, and the whole
sequence takes 25min integration in total.
We used the bright quasars 3C273, 2C279, or 3C345,
depending on availability during the observations.
For any pair of band and bandwidth, this sequence has simultaneous integrations 
which can be used to calibrate the phase offset and amplitude scale between bands.
The calibration observations took typically 45 min including the radio pointing and antenna slew.
These integrations were used for passband calibration as well.

An individual observation consisted of a 4-10 h track. The total observing time 
(after flagging tracks under bad weather) is about 230 h ($\sim 30$ tracks).
A typical track starts with radio pointing observations of a bright
quasar available at the time, then
observes a flux calibrator (e.g. a planet), and repeats the 25 min observing
cycle of gain calibrator ($\sim$ 5 min) and target (20 min including antenna slew for mosaic).
The passband/hybrid observations were performed at the middle of a track
when M51 is at a high elevation ($\sim 80$ deg).
At such a high elevation, each antenna slew between M51 and the calibrator takes
a considerable amount of time. Observing a passband calibrator at a lower elevation
avoids this loss.
The system temperature (Tsys) was measured every
gain calibrator cycle, and the atmospheric gain variation is corrected
real-time using Tsys. We observed 1153+495 as a gain calibrator.

The telescope pointings were corrected every 4 h during the night and
every 2 h during
daytime. The last $\sim 10$ tracks of the 30 total tracks also
included an additional optical pointing procedure
developed by \citet{cor10}.
The optical procedure can operate during daytime, as well as at night,
and a pointing correction was made every gain calibration cycle.
This method measures the offset between radio and optical pointing vectors
at the beginning of track (which is stable over periods much longer than the typical observation).
During the observing cycle of gain calibrator and target,
the pointing drift, typically several arcsec per hour, is adjusted
using a bright star close to the gain calibrator
using an optical camera. The overhead of the pointing adjustment is
less than 1 min.

\begin{figure}
\plotone{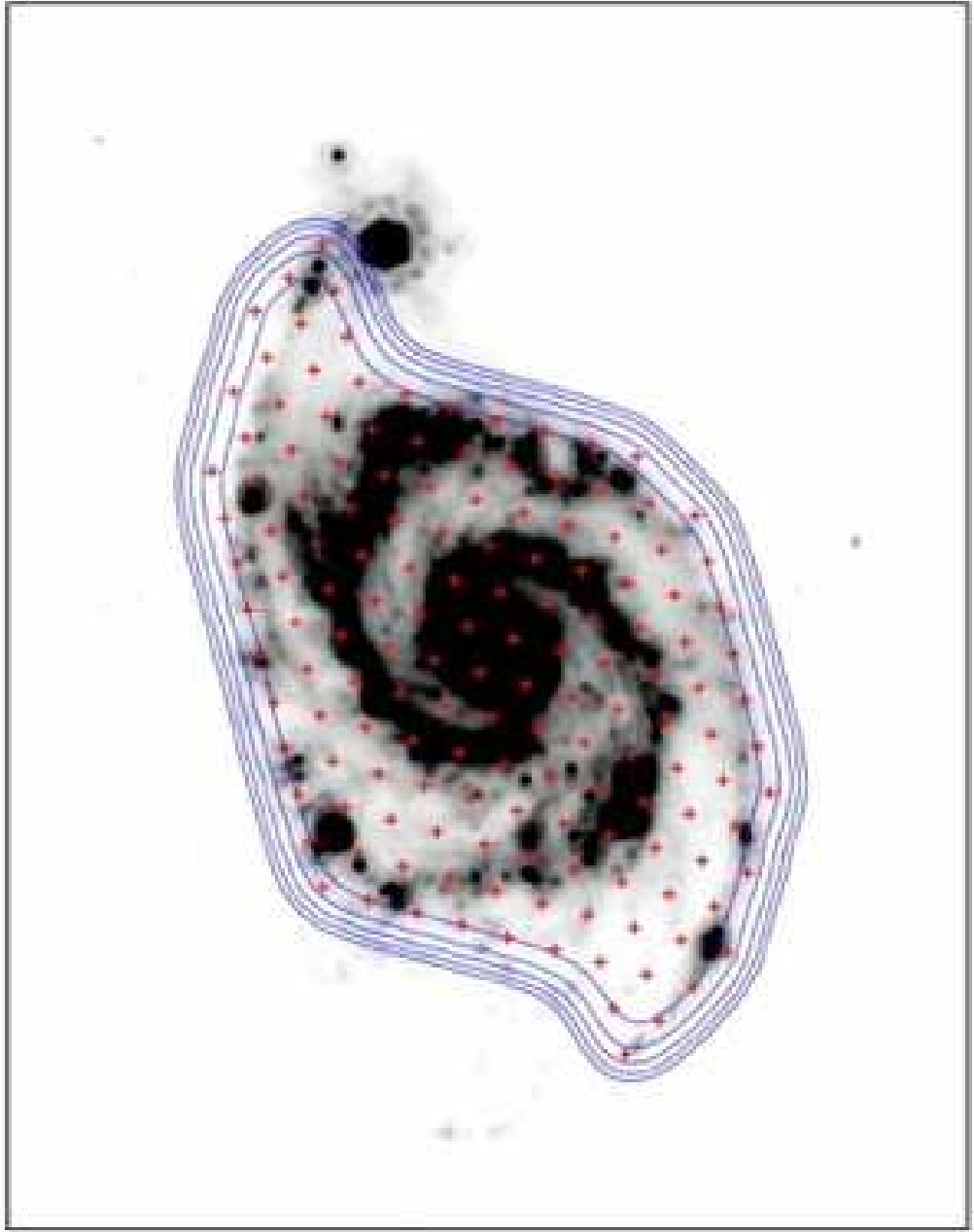}
\caption{Observed areas of M51 over the Spitzer 24$\mu$m image of M51. Red crosses: 151 pointing positions of CARMA observations. Blue contours:  sensitivities (RMS noise) of 120, 140, 160, 180, and 200 \% with respect to the central maximum sensitivity. Most part over the galactic disk has a uniform sensitivity. Black box: NRO45 coverage.
 \label{fig:pointing}}
\end{figure}

\placefigure{fig:pointing}

\begin{figure}
\plotone{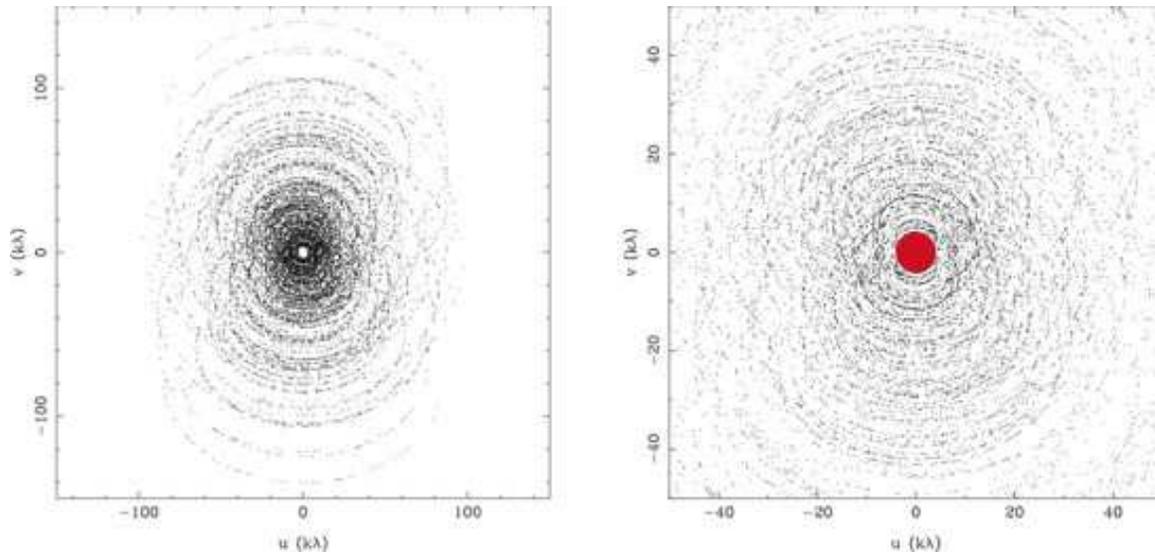}
\caption{The $uv$ coverage at the central pointing in unit of kilo-lambda. {\it Left: } CARMA $uv$ coverage. {\it Right:} the central region of CARMA (black) and NRO45 (red) $uv$ coverages.
 \label{fig:uvcov}}
\end{figure}

We mosaiced the entire $6 \arcmin .0 \times 8\arcmin .4$ disk of M51,
with the disk defined by optical images and shown in Figure \ref{fig:pointing}, 
in 151 pointings
with Nyquist sampling of the 10m antenna beam (FWHM of 1 arcmin for
the 115GHz CO J=1-0 line). Ideally, every pointing position would be observed
every M51 observing cycle ($\sim$ 20 min duration)
to maintain a uniform data quality and $uv$ coverage across the mosaiced area.
However, the overhead for slewing is significant for the large mosaic.
It is as long as 6 sec per slew, and about 15 min total for 151 pointings.
We therefore observed every third pointing (total $\sim50$ pointings)
in each observation cycle to reduce the overhead. Three consecutive cycles cover
all 151 pointings. Each track started from a pointing randomly chosen
from the table of the 151 pointings, which helps the uniform data quality among
pointings.
The resultant CARMA $uv$ coverage is very similar at all pointings, and
an example of the $uv$ coverage at the central pointing is in Figure \ref{fig:uvcov}.

\placefigure{fig:uvcov}

The primary flux calibrators, Uranus, Neptune, and MWC349,
were observed in most tracks. We monitored the flux of gain calibrator
1153+495 every month over the course of the observations. The flux of 1153+495 varied slowly between
0.7 and 1.3 Jy. The CARMA observatory is separately monitoring the flux variations
of common passband calibrators, and our flux measurements are
consistent with the observatory values.

\subsection{Calibration} \label{sec:redcarma}

The data were reduced and calibrated using the Multichannel Image
Reconstruction, Image Analysis, and Display (MIRIAD) software package
\citep{sau95}. We developed additional commands/tasks to investigate
and to reduce the large amount of data effectively, and to combine
interferometer and single-dish data.

The initial set of calibrations are the required routines for most CARMA data reduction.
First, we flag the data with problems such as antenna shadowing and
bad Tsys measurements. Second, we apply the correction for variation of optical fiber cable length,
namely line length correction.
CARMA is a heterogeneous array of two types of antennas (i.e., 6m and
10m), and 
the optical fiber cables that connect the antennas to the control building are
mounted differently for the 10m and 6m dishes. The time variations of the cable lengths due to thermal
expansion are therefore different,
which results in phase wraps in the baselines between 6m and 10m antennas.
The changes of the cable lengths were monitored to an accuracy of 0.1 pico-second
by sending signals from the control building and measuring their round-trip travel time.
The changes are stored in MIRAD data and are
used for the line-length correction. Third, we smooth the spectra with the Hanning window function to reduce the
high side-lobes in raw spectra from the digital correlators.
The spectral resolution is lowered by a factor of 2 and becomes 1.954 MHz
(5.08 km/s at the CO($J=1-0$) frequency).

Calibrations for passband and hybrid correlator configuration were made using
the sequence of hybrid configuration observations described in \S \ref{sec:obscarma}.
We first separate 500 MHz and 62 MHz integrations from the sequence 
and make two MIRIAD data sets containing only 500MHz data or 62 MHz data.
These data sets are used to derive and apply passbands. The passband calibration
removes the phase and amplitude offsets among Band 1, 2, and 3
in the 500 and 62 MHz modes. An offset/passband calibrator is significantly
detected even in 10 sec integration both in the 62 MHz and in the 500 MHz mode.
We derive the phase offset and amplitude scale between the 500 MHz and 62 MHz modes
by comparing the visibilities from the two modes on the 10 sec integration basis,
and averaging them over time to derive single values for the phase offset and amplitude
scale. We applied these calibrations to the entire track,
which removes the phase and
amplitude offsets between gain calibrator and target integrations.
Errors of the hybrid calibration are small compared to the other errors
and are only a few percent in amplitude and a few degrees in phase.

The last set of calibrations includes the standard phase calibrations
to compensate for atmospheric and instrumental phase drifts. We did
not use the gain calibrator integrations with large phase 
scatters (due to bad weather) and flagged the target integrations
in the cycles immediately before and after the bad gain data. The absolute fluxes
of the gain calibrator were measured monthly against a planet (\S \ref{sec:obscarma})
and were applied to target data.

The resulting $1\sigma$ noise level of the CARMA data is 27 mJy/beam in
each $10\kmps$ channel.

\section{Nobeyama Radio Observatory 45m Telescope}\label{sec:nro45}

\subsection{Observations} \label{sec:obsnro45}

We obtained total power and short spacing data with the 5x5-Beam Array
Receiver System \citep[BEARS; ][]{sun00} on the Nobeyama Radio
Observatory 45m telescope (NRO45).
The FWHM of the NRO45 beam is $15\arcsec$ at 115 GHz.
We configured the digital spectrometer \citep{sor00} to 512 MHz
bandwidth at 500 kHz channel resolution.
This is wide enough to cover the entire M51 system (both NGC 5194 and 5195).
Hanning smoothing was applied to reduce the side-lobe in channel,
and therefore, the resolution of raw data is 1 MHz.

We scanned M51 in the RA and DEC directions using the On-The-Fly (OTF) mapping
technique \citep{man07,saw08}. We integrated OFF positions around the galaxy before
and after each $\sim 1$ min OTF scan. A scan starts from an emission-free position at
one side of the galaxy and ends on another emission-free position at the other side.
Spectra are read-out every 0.1 second interval during the scan.
The receiver array was rotated by $7 \deg$ with respect to the
scan directions, so that the 25 beams draw a regular stripe with a $5\arcsec$ separation.
In combining the RA and DEC scans, the raw data form a 
lattice with $5\arcsec$ spacing.
This fine sampling, with respect to the beam size of $15\arcsec$, is necessary
in reproducing the $uv$ data up to the 45m baseline (i.e. the diameter of NRO45),
since we need the Nyquist sampling ($5.96\arcsec$) of $\lambda_{\rm CO}/D=11.92\arcsec$,
where $\lambda_{\rm CO}$ is the wavelength ($=2.6\,\rm mm$) and $D$ is
the antenna diameter \citep{man07}.
If the sampling is coarser than $5.96\arcsec$, the aliasing effect
in the Fourier space contaminates even shorter baseline data significantly.
For example, if the sampling spacing is only $10.3\arcsec$ \citep[i.e., typical sampling in
past NRO45 observations;  ][]{kun07}, the $uv$ data down to
the $\sim 7\,\rm m$ baseline is contaminated (Figure \ref{fig:uvnro45}),
and cannot be combined with the interferometer data.

\placefigure{fig:uvnro45}

The typical system temperature in double side band was $\sim320$ K.
The pointing of the telescope was checked every $\sim 45$ min and was
accurate to
within $\sim 2$-$3\arcsec$. BEARS is an array of double-side band (DSB) receivers
and provides the antenna temperature $T_a^*(\rm DSB)$ in DSB.
The upper/lower side band ratio, namely the scaling factor, was measured by observing
Orion IRC2 using both BEARS and the single-side band (SSB) receiver S100
and taking the ratios of the two measurements.
The error in the measurements is a few percent.
The total observing time under good weather conditions is about 50 hours.

\subsection{Calibration}\label{sec:nro45red}

The ON/OFF calibration to account for the sky background level was
applied after the observations. We interpolated between two 
OFF-sky integrations before and after each OTF scan ($\sim1$ minute long),
which reduced non-linear swells in the spectral baselines significantly.
We used the {\it NOSTAR} data reduction package developed at the Nobeyama
Radio Observatory \citep{saw08}, converted the flux scale from $T_a^*(\rm DSB)$ of BEARS
to $T_a^*(\rm SSB)$ of S100, subtracted linear spectral baselines,
and flagged bad integrations.

The 5" lattice of data from the observations was re-gridded with a spheroidal smoothing function,
resulting in a final resolution of $19.7\arcsec$.
We used the grid size of $5.96\arcsec$, which is the Nyquist sampling of the 45 m spacing
in Fourier space; this pixel scale is necessary to prevent artifacts from
the aliasing effect (\S \ref{sec:obsnro45}).

We made maps of the RA and DEC scans separately. The two maps were co-added
after subtracting spatial baselines in each scan direction to reduce systematic errors
in the scan direction.
Note that for OTF mapping, the sharing of an OFF among many ON scans may introduce noise
correlations, primarily at small spatial frequencies in the Fourier space.
 \citet{eme88} reduced such correlated noise using the basket-weave method,
 which down-weights the data at small spatial frequencies in the scan directions
 when the RA and DEC maps are added.
 We compared the spatial-baseline subtraction and basket-weave methods,
 and found that both diminish the large-scale noise well. The difference was subtle, but
 the former gave a slightly smaller RMS noise, and thus, we decided to use the
 spatial-baseline method.

The antenna temperature $T_a^*(\rm SSB)$ was converted to the main beam temperature
$T_{\rm mb}$, using the main beam efficiency of $\eta_{\rm mb} = 0.4$ and
$T_{\rm mb} = T_a^*(\rm SSB)/\eta_{\rm mb}$.

The flux of the final NRO45 map is consistent with most previous measurements
within a typical error of millimeter-wave measurements (10-20\%).
It is compared with four other results: an image from the National Radio Astronomy
Observatory 12 m telescope \citep[NRAO12; ][]{hel03},
two previous measurements at NRO45 \citep{nak94, mat99},
and our new CARMA data (\S \ref{sec:carma}).
The fluxes from \citet{hel03}, \citet{mat99}, and the new CARMA observations are
94\%, 95\%, and 93\% of that of the new NRO45 map, respectively.
For the comparisons, we re-sampled the new map to match the area coverage
of the other maps.
For the comparison with CARMA, the CARMA {\it uv}-distribution is generated
from the new NRO45 map (as discussed in \S \ref{sec:nrouv}, but for Hatcreek, OVRO,
and CARMA primary beams), and the positive fluxes (above about $4\sigma$)
in the dirty maps are compared to measure the flux ratio.
We used a Gaussian taper (FWHM=$20\arcsec$) to make the dirty maps,
which roughly reproduces the weight distribution of the NRO45 data.

Only the map of \citet[][distributed through \citet{kun07}]{nak94} shows a significant discrepancy:
a factor of 1.82 higher total flux than the new NRO45 map.
We attribute this discrepancy to an error in the old map, since all other
measurements are consistent.
Among these measurements, we decided to rely on the CARMA flux
because we had the best understanding of the process of flux calibration,
and because it is based on multiple flux calibrations over the duration of the observations.
We scaled the flux of the NRO45 map to match the CARMA flux
(i.e., multiplied 0.93).

The $1\sigma$ noise level of NRO45 data is 14.7 mK in $T_a^*(\rm SSB)$, 36.7 mK in
$T_{\rm mb}$, and 155 $\rm mJy/beam$ in $10\kmps$ channel.


\section{Imaging Heterogeneous-Array Mosaic Data}\label{sec:imaging}

We use MIRIAD for joint-deconvolution of multi-pointing CARMA and NRO45 data.
The method and algorithm for mosaic data with a homogeneous array are
described in \citet{sau96}. Our imaging involves two additional
complications: a
heterogeneous array, and combinations with single-dish data, as well as mosaicing. 
We describe the essence of joint-deconvolution using MIRIAD, with
an emphasis on the case of CARMA and NRO45.

Two points are of particular importance: the treatment of different primary beam
patterns, and the weights of the data from the different primary beam patterns
and from the single-dish. Here, we illustrate these two points, and define our
notations.

Correction for primary beam attenuation is simple for a homogeneous array.
All antennas have the same primary beam
pattern $P(l,m)$, and the primary beam correction is
\begin{equation}
I(l,m) = \frac{\bar{I}(l,m)}{P(l,m)},
\label{eq:pb1}
\end{equation}
where the primary-beam corrected image is denoted $I$ and the
uncorrected image is denoted $\bar{I}$. The sky
coordinates are $(l,m)$.
The uncorrected image $\bar{I}$ has two advantages:
the synthesized beam $\bar{B}$ (i.e., point spread function, PSF) and
noise level are position-invariant,
which simplifies the process of deconvolution (\S \ref{sec:homsin}).

For a heterogeneous array, the differences between primary beam patterns
have to be taken into account. For example, CARMA has three
baseline types (i.e., antenna pairs), which result in three primary
beam patterns -- called "H" for Hatcreek (6m-6m dish pair), "O" for OVRO (10m-10m),
and "C" for CARMA baseline types (6m-10m).
Using appropriate weights $W_{\rm H}$, $W_{\rm O}$, and $W_{\rm C}$
(\S \ref{sec:weighting}), the images from "O", "H", and "C" baselines can
be added as
\begin{eqnarray}
I(l,m) &=&
W_{\rm H} \frac{\bar{I}_{H}}{P_{\rm H}} +
W_{\rm O} \frac{\bar{I}_{O}}{P_{\rm O}} +
W_{\rm C} \frac{\bar{I}_{C}}{P_{\rm C}} \nonumber \\
&=&
W_{\rm H} I_{H} +
W_{\rm O} I_{O} +
W_{\rm C} I_{C}.
\label{eq:pb3}
\end{eqnarray}
The weight $W$ is a function of position $(l,m)$. The co-added image has 
been corrected for primary beam attenuation. In the co-added plane,
the synthesized beam pattern $B$ and noise level are position-variant,
which complicates the deconvolution.


\subsection{Homogeneous Array, Single-Pointing Data} \label{sec:homsin}
Traditionally, the imaging of interferometer data has been performed as follows.
A set of {\it one} dirty map $\bar{I}^{\rm dm}(l,m)$ and {\it one} synthesized
beam pattern $\bar{B}(l,m)$ is made from visibilities.
The dirty map $\bar{I}^{\rm dm}$ is deconvolved with $\bar{B}$.
For example, the deconvolution scheme CLEAN replaces the pattern
$\bar{B}(l-l_0,m-m_0)$,
centered at an emission peak at ($l_0$, $m_0$), with an ellipsoidal Gaussian
to reduce the sidelobes of  $\bar{B}$.
CLEAN usually runs in the $\bar{I}^{\rm dm}$ domain;
the synthesized beam  $\bar{B}$ and noise level $\sigma$ are
position-invariant and their treatments are simple.
The CLEANed image $\bar{I}^{\rm mp}$ is corrected for primary beam attenuation
(eq. (\ref{eq:pb1})), providing the final map $I^{\rm mp}$.
We note again that primary beam uncorrected and corrected images (of any kind)
are differentiated with "bar" (e.g., $\bar{I}^{\rm dm}$ vs. $I^{\rm dm}$ and
$\bar{I}^{\rm mp}$ vs. $I^{\rm mp}$).

The deconvolution is also possible in the $I^{\rm dm}$ domain.
The synthesized beam pattern $B$ and noise level are {\it not}
position-invariant. Thus, we define a position-variant synthesized beam pattern,
\begin{equation}
B(l,m; l_0,m_0) = \frac{\bar{B}(l-l_0,m-m_0)}{P(l,m)}, \label{eq:synbm}
\label{eq:bmpb}
\end{equation}
centered at $(l_0,m_0)$, and a position-variant noise level $\sigma/P(l,m)$.
Emission peaks are searched on a basis of signal-to-noise ratio.
In MIRIAD, a set of primary-beam corrected $I$ and uncorrected $\bar{B}$ is
calculated from visibilities, and the command "mossdi" (i.e., CLEAN)
calculates $B$ with eq. (\ref{eq:synbm}) at peak position $(l_0,m_0)$.


\subsection{Heterogeneous Array, Single Pointing Data} \label{sec:hetero1}

The deconvolution in the image domain is applicable to heterogeneous array data.
The joint dirty image $I^{\rm dm}$ is defined as a linear
summation of three dirty maps $I^{\rm dm}_{\rm H}$, $I^{\rm dm}_{\rm O}$, and
$I^{\rm dm}_{\rm C}$ (eq. (\ref{eq:pb3})).
The corresponding synthesized beam $B$ is also a linear summation with the
same weights,
\begin{eqnarray}
B(l,m; l_0,m_0) &=& W_{\rm H}(l,m) \frac{\bar{B}_{H}(l-l_0, m-m_0)}{P_{H}(l,m)} \nonumber \\
&+& W_{\rm O}(l,m) \frac{\bar{B}_{O}(l-l_0, m-m_0)}{P_{O}(l,m)} \nonumber \\
&+& W_{\rm C}(l,m) \frac{\bar{B}_{C}(l-l_0, m-m_0)}{P_{C}(l,m)}.
\label{eq:bm3}
\end{eqnarray}

The MIRIAD command "invert" with the mosaic option outputs a set of
{\it one} joint dirty map $I^{\rm dm}$ (primary beam corrected) and {\it three}
synthesized beams
$\bar{B}_{\rm H}$, $\bar{B}_{\rm O}$, and $\bar{B}_{\rm C}$ (uncorrected).
The command "mossdi" finds a peak emission in $I^{\rm dm}$,
and calculates  $B$ at its position with eq. (\ref{eq:bm3}).

In the case of a heterogeneous array, such as CARMA, the primary beam correction
always needs to be applied to the dirty map.
Thus, even for single-pointing observations, we always use "options=mosaic" for "invert".


\subsection{Heterogeneous Array, Mosaic Data}\label{sec:hetmos}

The deconvolution of mosaic data with a heterogeneous array is
a further extension of the same procedure.
Eq. (\ref{eq:pb3}) is extended as
\begin{equation}
I (l,m) = \sum_{b,p} W_{b,p} \frac{\bar{I}_{b,p}}{P_{b,p}} = \sum_{b,p} W_{b,p} I_{b,p}\\
\label{eq:pbm}
\end{equation}
where the summation is taken for all baseline types $b$ and pointings $p$.
$W_{b,p}$ is a weight for $b$ and $p$.
In practice, $P$ is truncated at some radius, and only a subset of
pointings contribute to a given position.

The joint synthesized beam is defined as in eq. (\ref{eq:bm3}),
but includes all pointings.
In the case of the CARMA M51 observations, the command "invert" with "option=mosaic"
outputs {\it one} joint dirty map
and 453 synthesized beams (= 3 baseline types $\times$ 151 pointings).
A joint synthesized beam $B$ is calculated with the 453 synthesized beams
for every emission peak in $I^{\rm dm}$.

The spatial resolution is calculated by taking a weighted average of all
453 synthesized beams using $W_b$ ($b=$H, O, C) and by fiting a Gaussian.
In theory, the sizes of the synthesized beams are different among the pointings,
since the $uv$ coverage is not exactly the same for all of the pointings.
In practice, we designed the observations to provide uniform $uv$ coverage
for all pointings (\S \ref{sec:obscarma}). We therefore adopt a single
beam size over the whole mosaic.


\subsection{Weighting}\label{sec:weighting}

The noise level is position-dependent, $\sigma /P(l,m)$, in the image domain.
Therefore, the weights $W$ are defined as
\begin{equation}
W_b(l,m) \propto \left( \frac{P_b(l,m)}{\sigma}\right)^2
\label{eq:w}
\end{equation}
for $b=$ H, O, and C, and are normalized as $W_{\rm H} + W_{\rm O} + W_{\rm C} = 1$
at each position $(l,m)$. The theoretical noise $\sigma$ depends on baseline type $b$,
and is the same as the imaging sensitivity $\Delta S^{\rm i}$ discussed below.


\subsubsection{Thermal Noise and Its Coefficient}\label{sec:noise}

Two sensitivities, fringe sensitivity $\Delta S^{\rm f}$ and imaging sensitivity
$\Delta S^{\rm i}$ [$\equiv \sigma$],
are important \citep[][see their section 9]{tay99}. The fringe sensitivity is a sensitivity per visibility.
The theoretical sensitivity $S^{\rm f}$ for each visibility is calculated with the system
temperature $T_{\rm sys}$, bandwidth $B$, and integration time
of the visibility $t_{\rm vis}$ as
\begin{equation}
\Delta S^{\rm f} = C_{ij} \sqrt{\frac{ T_{{\rm sys}, i} T_{{\rm sys}, j}}{B \cdot t_{\rm vis}}},
\label{eq:dsk}
\end{equation}
where
\begin{equation}
C_{ij} = \frac{2 k_{\rm B}}{ \sqrt{(\eta_{a, i} A_i)(\eta_{a, j} A_j)} } \frac{1}{\sqrt{2} \eta_q}.
\label{eq:cij}
\end{equation}
The aperture efficiency $\eta_a$ and collecting area $A$ of antennas
$i$ and $j$ have a relation with the beam solid angle $\Omega_{\rm A}$
given by $1/(\eta_a A)= \Omega_{\rm A} / \lambda^2$.
The Boltzman constant is $k_{\rm B}$ and 
the last term $1/\sqrt{2}\eta_q$ is due to the backend (i.e., digitizer and correlator),
and $\eta_q$ is the quantum efficiency \citep{roh00}.
$C_{ij}$ is approximated as a constant for a homogeneous array, since the parameters
are very similar for all antennas.
In the case of a heterogeneous array, $C_{ij}$ depends on baseline type.
Parameters are listed in Table \ref{tab:ant}.

The imaging sensitivity is a root-mean-square (RMS) noise in a final image, and
depends on control parameters (see Appendix \ref{sec:app}; e.g., natural and uniform weighting).
If the natural weighting is employed, the imaging sensitivity is simply a statistical summation
of fringe sensitivities, $1/(\Delta S^{\rm i})^2 = \Sigma_k 1/(\Delta S^{\rm f})^2$. For a homogeneous array
($C_{ij}$ is constant), it is
\begin{equation}
\Delta S^{\rm i} [\equiv \sigma] = C_{ij}  \sqrt{\frac{ T_{{\rm sys}, i} T_{{\rm sys}, j}}{B \cdot t_{\rm tot} }}, \label{eq:senim}
\end{equation}
assuming that $T_{\rm sys}$ is a constant during observations.
The total integration time is $t_{\rm tot} = N_{\rm vis} t_{\rm vis}$, where
$N_{\rm vis}$ is the number of visibilities.



\section{The Combination of NRO45 with CARMA} \label{sec:comb}

\begin{figure}
\epsscale{1.00}
\plotone{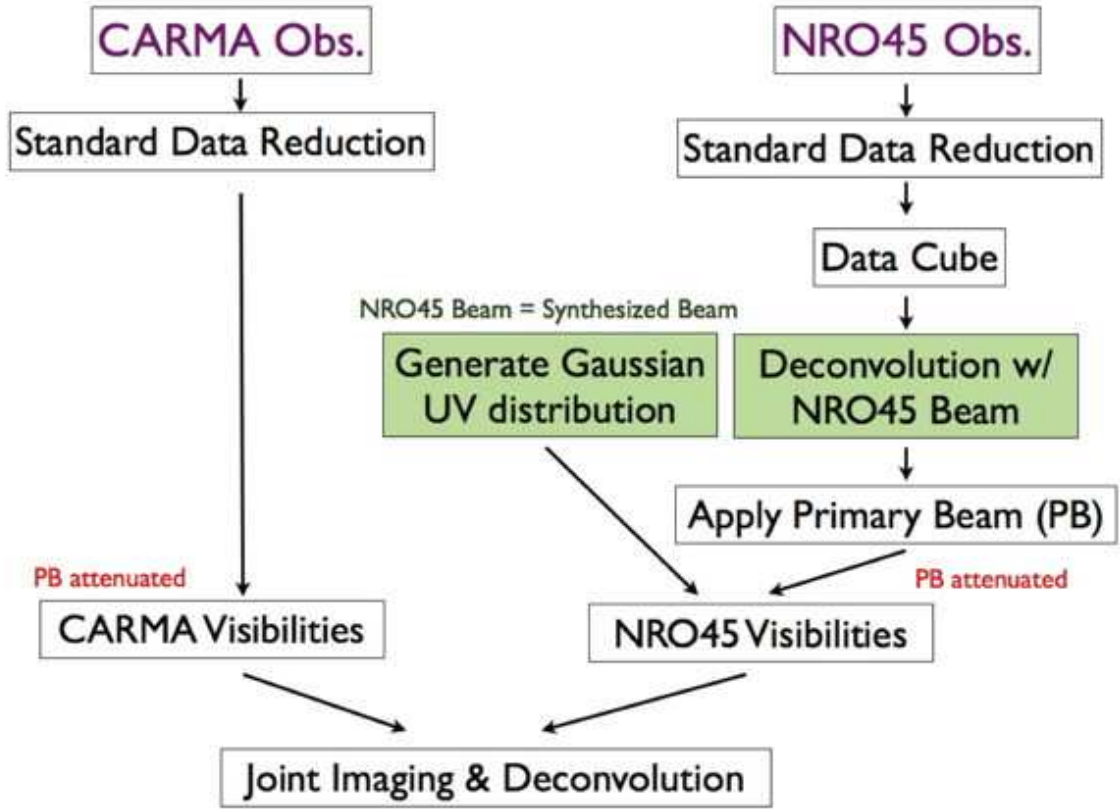}
\caption{Flow chart of the process of combination. The NRO45 cube is deconvolved with the NRO45 beam, multiplied with an arbitrarily selected primary beam, and Fourier-transformed into $uv$-space. The transformed data are re-sampled with a Gaussian $uv$ distribution to produce NRO45 visibilities. The weight of the NRO45 visibilities, with respect to CARMA, are determined based on the RMS noise of the NRO45 cube. 
 \label{fig:combflow}}
\end{figure}

The NRO45 image is converted to visibilities and combined with CARMA data
in $uv$ space. Here we discuss four steps for combination:
(1) generating visibilities from the single-dish image,
(2) the calculation of the weights of the single-dish visibilities, in the same form
as interferometer visibilities,
(3) the determination of synthesized beam size, and
(4) an imaging/deconvolution scheme.
A flow chart of the procedure is in Figure \ref{fig:combflow}.

\placefigure{fig:combflow}


\subsection{Converting the NRO45 Map to Visibilities}\label{sec:nrouv}

To produce NRO45 visibilities, we first deconvolve a NRO45 map with
a NRO45 point spread function (PSF),
multiply a dummy primary beam, generate a Gaussian visibility
distribution, and calculate the amplitude and phase of the visibilities
from the deconvolved, primary-beam applied NRO45 map
(Figure \ref{fig:combflow}).
The following sections describe these steps.

One limitation arises from the current software, though
it should be easily modified in future software development.
NRO45 visibilities must have the same form as those of interferometers,
and therefore, a dummy primary beam needs to be applied to the NRO45 map.

\subsubsection{Deconvolution with the NRO45 Beam}\label{sec:deconvnro45}

A NRO45 map is a convolution of a true emission distribution with a point spread
function (PSF). In the case of OTF mapping (\S \ref{sec:obsnro45}), the PSF is
not literally the NRO45 beam, but is a convolution of the NRO45 beam and
the spheroidal function which is used to re-grid the observed data to a map grid
(\S \ref{sec:nro45red}).
The intrinsic Gaussian FWHM of the NRO45 beam is  $15\arcsec$, and is
degraded to $19.7\arcsec$ after the re-gridding. The NRO45 map needs to
be de-convolved with this PSF.

\begin{figure}
\epsscale{0.80}
\plotone{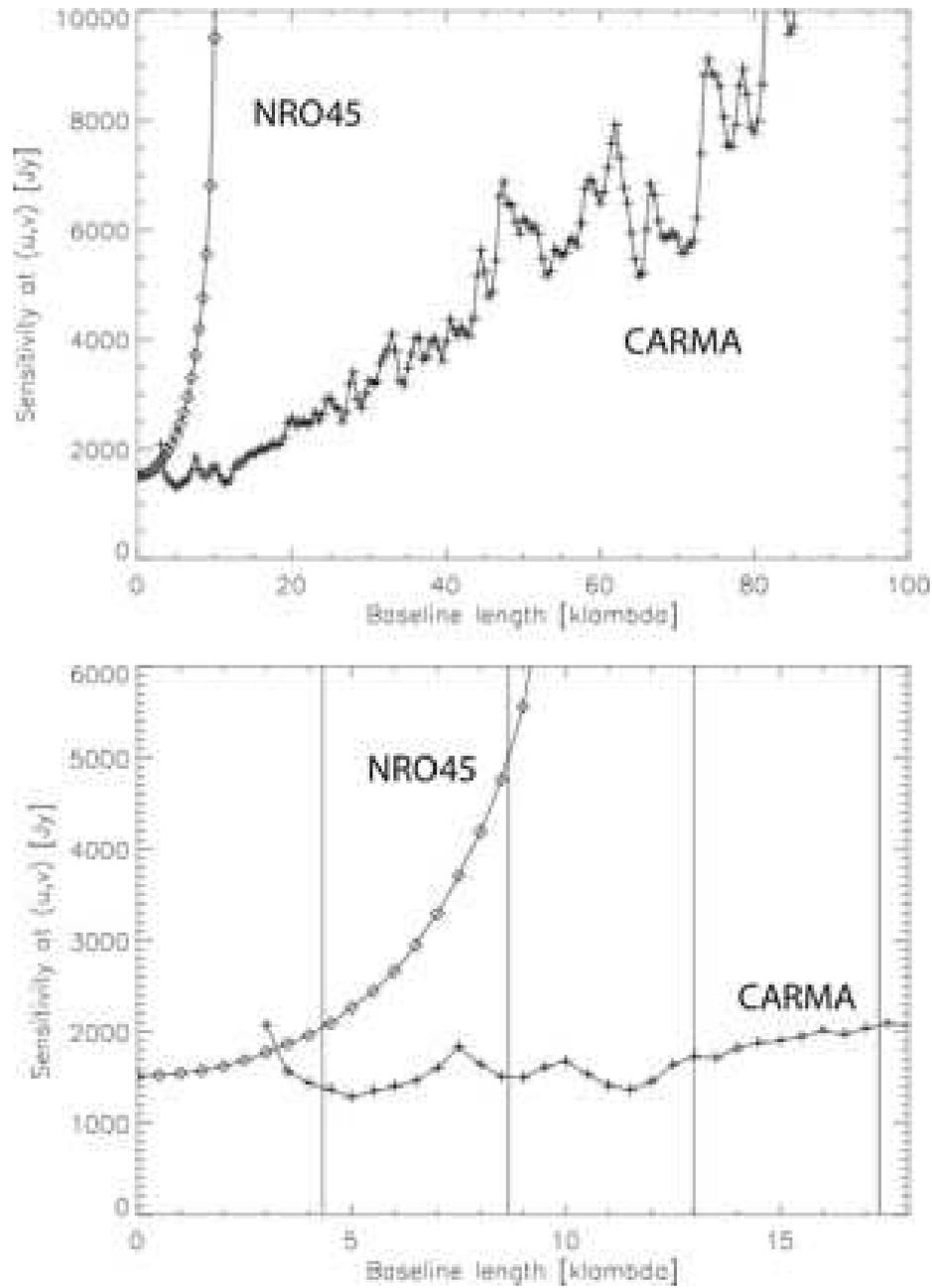}
\caption{Sensitivities as a function of baseline length for NRO45 (diamonds) and CARMA (crosses). The bottom panel is the same as the top panel, but only for short baselines. Vertical lines, from right to left, correspond to the antenna diameter of NRO45 (45m) and its three quarter, half, and quarter lengths.
 \label{fig:sen_data}}
 \end{figure}

Figure \ref{fig:sen_data} shows the sensitivity (noise) as a function of $uv$-distance (baseline length).
It has a dependence on the Fourier-transformed PSF  FT\{PSF\} as $\propto 1/\sqrt{\rm FT\{PSF\}}$
(see Appendix \ref{sec:senmatch}).
The standard deviation of FT\{PSF\} is $\sigma_{\rm F}=3.9 \,\rm k\lambda$
for a Gaussian PSF with the FWHM of $19.7\arcsec$. Thus, the noise increases
significantly beyond 4-6 k$\lambda$ (i.e. $\sqrt{2} \sigma_{\rm F}$).
Figure \ref{fig:sen_data}  shows that the NRO45 sensitivity is comparable to
that of CARMA up to 4-6 k$\lambda$ and deviates beyond that.
With the resultant sensitivities, we decided to flag the data at  $>4 \rm k\lambda$.
The long baselines have negligible effects if we use only the weight based on sensitivity
\citep[i.e. {\it robust=+2}, ][]{bri95}, but could introduce an elevated error when $robust < +2$.  

CARMA and NRO45 are complementary in terms of $uv$-coverage and sensitivity
(Figure \ref{fig:uvcov} and  \ref{fig:sen_data}). \citet{kur09} suggested that the single-dish
diameter should be 1.7 times as large as the minimum baseline of interferometer data,
which is $\sim 18$ meters in our case. However,  we seem to need a 45m class telescope
to satisfy the sensitivity requirement within realistic observing time.
The sensitivity matching between NRO45 and CARMA data is discussed
in Appendix \ref{sec:senmatch}.

\subsubsection{Applying a Dummy Primary Beam}

The imaging tasks in MIRIAD assume that all visibilities are from interferometric observations,
and apply a primary beam correction in the process of imaging. Consequently, the NRO45
visibilities need to be attenuated by a pseudo primary beam pattern $P_{\rm N}$.
The choice of $P_{\rm N}$ is arbitrary, and we employ a Gaussian primary
beam with the FWHM of 2 arcmin. 
$P_{\rm N}$ is multiplied to the deconvolved NRO45 map at each of the 151 CARMA
pointings separately. Since the map will be divided by $P_{\rm N}$ during the deconvolution,
 the choice of $P_{\rm N}$ does not affect the result.
 However, it is safer to use $P_{\rm N}$ at least twice as  large as the separation of the pointings,
 so that the entire field is covered at the Nyquist sampling (or over-sampling) rate.

We note that this multiplication of a primary beam in the image domain is equivalent to a convolution
in the Fourier domain. It smoothes the sensitivity distribution in $uv$-space, and therefore, the weight
discussed in \S \ref{sec:nrowei}. The size of the primary beam in Fourier space is only 1/6 of
that of the NRO45 beam. Therefore, this effect should be small and negligible.

\subsubsection{Generating a Gaussian Visibility Distribution}
The distribution of visibilities in $uv$ space should reproduce the NRO45 beam
(more precisely, the PSF in \S \ref{sec:nrouv}) as a synthesized beam in image space.
The Fourier transformation of a Gaussian PSF is a Gaussian. Therefore, visibilities
are distributed to produce a Gaussian density profile in $uv$ space.
The size of the Gaussian distribution is set to reproduce the beam size of $19.7\arcsec$.
We manually add a visibility at ($u$, $v$) = (0,0), so that the zero-spacing is always
included. The number of visibilities $N_{\rm vis}$ and integration time per visibility
$t_{\rm int}$ are control parameters, and are discussed in \S \ref{sec:nrowei}.

\subsubsection{Resampling}
From the Gaussian visibility distribution and the primary beam attenuated maps,
the visibility amplitudes and phases are derived, which gives the NRO45 visibilities.

\subsection{Theoretical Noise and Other Parameters}\label{sec:nrowei}

The relative weights of the CARMA and NRO45 visibilities are important for
proper combination. MIRIAD requires a weight (sensitivity) per individual
visibility for imaging, and we calculate the weight based on the RMS
noise of a NRO45 map. For the interferometer data (\S \ref{sec:noise}),
we start from the fringe sensitivity $S^{\rm f}$ and calculate the imaging
sensitivity $S^{\rm i}$ by summing up the $S^{\rm f}$s of all visibilities.
Here, we start from the RMS noise of a map
(i.e., $S^{\rm i}$) and determine $S^{\rm f}$ and its coefficient.

The theoretical noise of a single-dish map, in main beam temperature $T_{\rm mb}$, is
\begin{equation}
\Delta T_{\rm mb} = \frac{T_{\rm sys}}{\eta_q \eta_{\rm mb} \sqrt{B \cdot t_{\rm tot}}}, \label{eq:deltatmb}
\end{equation}
where $\eta_q$ and $\eta_{\rm mb}$ are the quantum efficiency of the spectrometer and
the main beam efficiency of the antenna, respectively.
$B$ and $t_{\rm tot}$ are the bandwidth and total integration time, respectively.
[Note that the contribution to the noise from the OFF position integrations should be negligible
in  OTF mapping \citep{saw08}].
The total integration time (per point) of the NRO45 map is derived from the RMS
noise in the map using this equation.

The imaging sensitivity, corresponding to eq. (\ref{eq:senim}), is calculated
by converting the unit of eq. (\ref{eq:deltatmb}) from Kelvin to Jy,
\begin{equation}
\Delta S^{\rm i} = \frac{2 k_{\rm B}}{\eta_a A} \frac{T_{\rm sys}}{\eta_q \eta_{\rm mb} \sqrt{B \cdot t_{\rm tot}}}.\label{eq:nro45isen}
\end{equation}
Comparing with eq. (\ref{eq:senim}), we obtain
\begin{equation}
C_{ij} = \frac{2 k_{\rm B}}{\eta_{\rm mb} \eta_a A } \frac{1}{\eta_q}.
\label{eq:nrocij}
\end{equation}

The fringe sensitivity per visibility should be
\begin{equation}
\Delta S^{\rm f} = C_{ij} \frac{T_{\rm sys}}{\sqrt{B\cdot t_{\rm vis}}}, \label{eq:nro45fsen}
\end{equation}
where the integration time per visibility is $t_{\rm vis} = t_{\rm tot} / N_{\rm vis}$
and $N_{\rm vis}$ is the number of visibilities. The $t_{\rm vis}$
value should be set (arbitrary)
to a small number, so that $N_{\rm vis}$ becomes large enough to fill the $uv$ space.
We set $t_{\rm vis} = 0.01 \rm \, sec$ and $N_{\rm vis} = 42075$.

Conceptually, we can understand the meaning of the NRO45 visibilities by comparing
the definition of fringe sensitivities (eqs. \ref{eq:dsk} and \ref{eq:nro45fsen}).
They are the ones observed virtually with two identical NRO45 antennas.
The two antennas can physically overlap (in our virtual observations), so
they can provide $uv$ coverage down to zero spacing. The beam shape of the NRO45
dish plays the role of synthesized beam, but not primary beam.
The primary beam shape is arbitrarily defined by $P_{\rm N}$ -- if we seek a meaning,
it corresponds to the beam shape of small patches within the NRO45 dishes.

There is one caveat when this weighting method is applied with the current version of MIRIAD.
MIRIAD is designed for an array with the same backend for all visibilities,
and therefore, it neglects the $1/\sqrt{2}\eta_q$ term from $C_{ij}$ (eq. \ref{eq:cij}).
It defines an alternative parameter,
\begin{equation}
{\rm JYPERK} = \frac{2 k_{\rm B}}{ \sqrt{(\eta_{a, i} A_i)(\eta_{a, j} A_j)} },
\label{eq:jyperk}
\end{equation}
which is stored in data header.
The weights ($\Delta S^{\rm f}$) are calculated with JYPERK, instead of  $C_{ij}$,
and do not take into account the backend.
In combining CARMA data with single-dish data,
we can overwrite JYPERK in CARMA data with $C_{ij}$, or define JYPERK
for single-dish (NRO45), as
\begin{equation}
{\rm JYPERK} = \frac{2 \sqrt{2} k_{\rm B}}{\eta_{\rm mb} \eta_a A } \left( \frac{\eta_{q, \rm CARMA}}{\eta_{q, \rm NRO45}} \right).
\label{eq:nrojyperk}
\end{equation}
Parameters are listed in Table \ref{tab:ant}.


\subsection{Synthesized Beam Size} \label{sec:synbm}
The deconvolution process (e.g., CLEAN) replaces a synthesized beam
with a convolution beam (typically a Gaussian).
We determine the convolution beam size so that its beam solid angle matches
that of the synthesized beam. Theoretically, the beam solid angle is an integration of
a beam response function over $4\pi$ steradians.
In principle, we could calculate it by integrating a synthesized beam image
or by taking the weight of the zero-spacing data (Appendix \ref{sec:solidangle}).
These methods worked reasonably well, but showed some error, introduced perhaps
by the limited size of the beam image (not over $4\pi$ steradians).
In practice, we found that the following method provides better flux conservation:
we calculate the total fluxes of the galaxy with the single-dish map and with
the dirty image (with an unknown beam area as a free parameter), and find the
beam area that equalizes these total fluxes.
The position angle and axis ratio of the beam is derived by a Gaussian
fitting to the synthesized beam. The Gaussian is linearly scaled to reproduce
the beam area from the flux comparison.

If the solid angles do not match, the total flux is not conserved in the final deconvolved
map (e.g., CLEANed map). The CLEANed map has two emission components,
deconvolved emission and noise/residual emission,
and they have their own units of flux, {\it Jy/(convolution beam)} and {\it Jy/(synthesized beam)},
respectively.
Therefore, the convolution beam smaller than the synthesized beam elevates
the flux of the residual emission, while a larger beam reduces it.
The error becomes particularly problematic for an object with extended, low-flux emission
(such as galaxies), which are inherently missed in the deconvolution process, but exist in
the CLEANed map. The two units in the final map does not degrade an image quality
much in case of the CARMA and NRO45 image, as long as the two beam areas are the same,
because the synthesized beam is already similar to a Gaussian beam that we adopt
as a convolution beam.

We note that if the deconvolution procedure (such as CLEAN) can 'dig' all positive components
down to the zero flux level, the convolution beam could have any shape.
We also note that in case of pure interferometer observations, the beam solid angle is zero
(Appendix \ref{sec:solidangle}), and thus, this method cannot be applied.


\subsection{Joint Imaging and Deconvolution}

The procedure for imaging and deconvolution is the same as the one in \S \ref{sec:hetmos},
but we add the term $W_{\rm N} I_{\rm N}$ in eq. (\ref{eq:pbm}),
\begin{equation}
I(l,m) =
W_{\rm H} I_{\rm H} +
W_{\rm O} I_{\rm O} +
W_{\rm C} I_{\rm C} +
W_{\rm N} I_{\rm N},
\end{equation}
where $I_{\rm N}$ and $W_{\rm N}$ are the image and weight from the NRO45 visibilities, respectively.
$W_{\rm N}$ is calculated with the pseudo primary beam $P_{\rm N}$ (\S \ref{sec:nrouv})
and the theoretical noise $\sigma_{\rm N}$ ($=\Delta S^{\rm i}$) derived with eq. (\ref{eq:nro45isen}).

The number of NRO45 visibilities is a control parameter in this method (\S \ref{sec:nrowei}),
and thus, should not be involved in weighting. Instead, the weight should be
calculated solely based on the sensitivity. We use the theoretical noise for weighting.
If all visibilities have exactly the same fringe sensitivity, our weighting becomes
the same as the conventional "natural" weighting. The sensitivity-based weighting for interferometer data
was discussed in \citet{bri95}, and our method extends it to the combination with single-dish data.

\begin{figure}
\epsscale{1.00}
\plotone{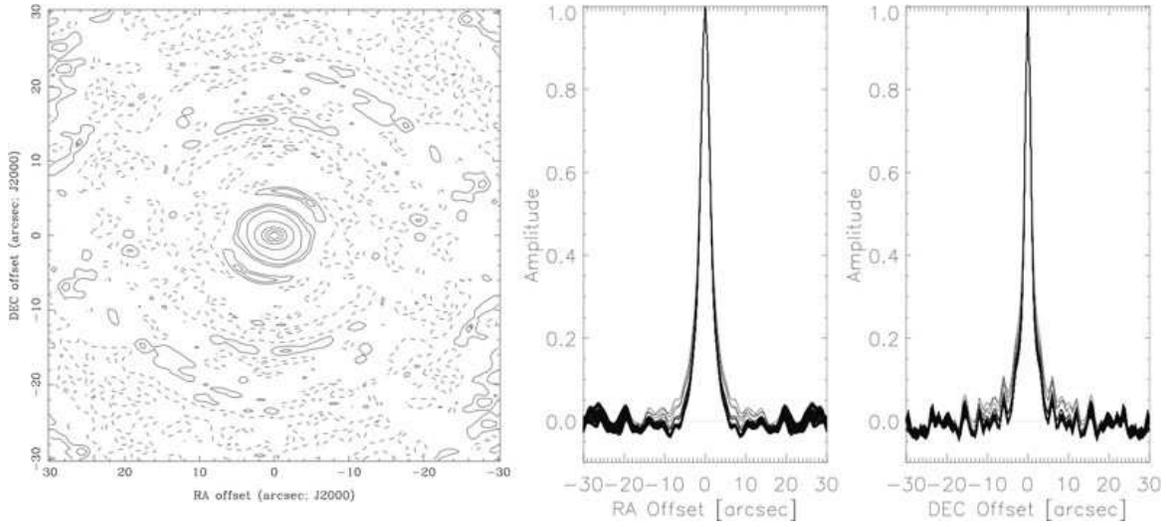}
\caption{Synthesized beam, i.e., the combination of the 4 baseline types (CARMA, HATCREEK, OVRO, and NRO45),  for {\it robust=+2}.
{\it Left:} Synthesized beam pattern. Contours are -4, -2, 2, 4, 10, 20, 40, 60, 80, 100 \% of the peak. Dashed lines are negative contours.
{\it Middle and right:} Slices of synthesized beams along RA and DEC directions, i.e. a superposition of all the beams at the 151 pointings. The synthesized beam patterns, i.e., $uv$ coverages, are very similar for all the pointings.
 \label{fig:synbeam}}
 \end{figure}

The robust weighting scheme suppresses the pixels of high natural weights in $uv$ space \citep{bri95} --
if the natural weight is lower than a threshold the weight is unchanged, but if it is higher, the weight
of the pixel is set to this threshold.
Our weighting scheme reproduces a natural weighting and works with the robust weighting scheme.
We made two data cubes with $robust = -2$ and $+2$. The resolution of the final combined
data cube with $robust =-2$ is $3.7\arcsec \times 2.9\arcsec$ (PA=$79\arcdeg$) and $5.08 \kmps$.
The RMS noise is 35 mJy/beam (i.e., 300 mK) in $10\kmps$ channel.
$robust =+2$ gives the resolution of $8.5\arcsec \times 7.3\arcsec$ (PA=$76\arcdeg$) and
$5.08 \kmps$, and the RMS noise of 52 mJy/beam (77 mK) in $10\kmps$ channel.
Figure \ref{fig:synbeam} shows a synthesized beam pattern for $robust =+2$.
Both cubes have the  same total luminosity when the synthesized beam sizes are determined as in \S \ref{sec:synbm}.

\placefigure{fig:synbeam}


\section{Integrated Intensity Map and Image Fidelity}\label{sec:fidelity}

\begin{figure}
\epsscale{.80}
\plotone{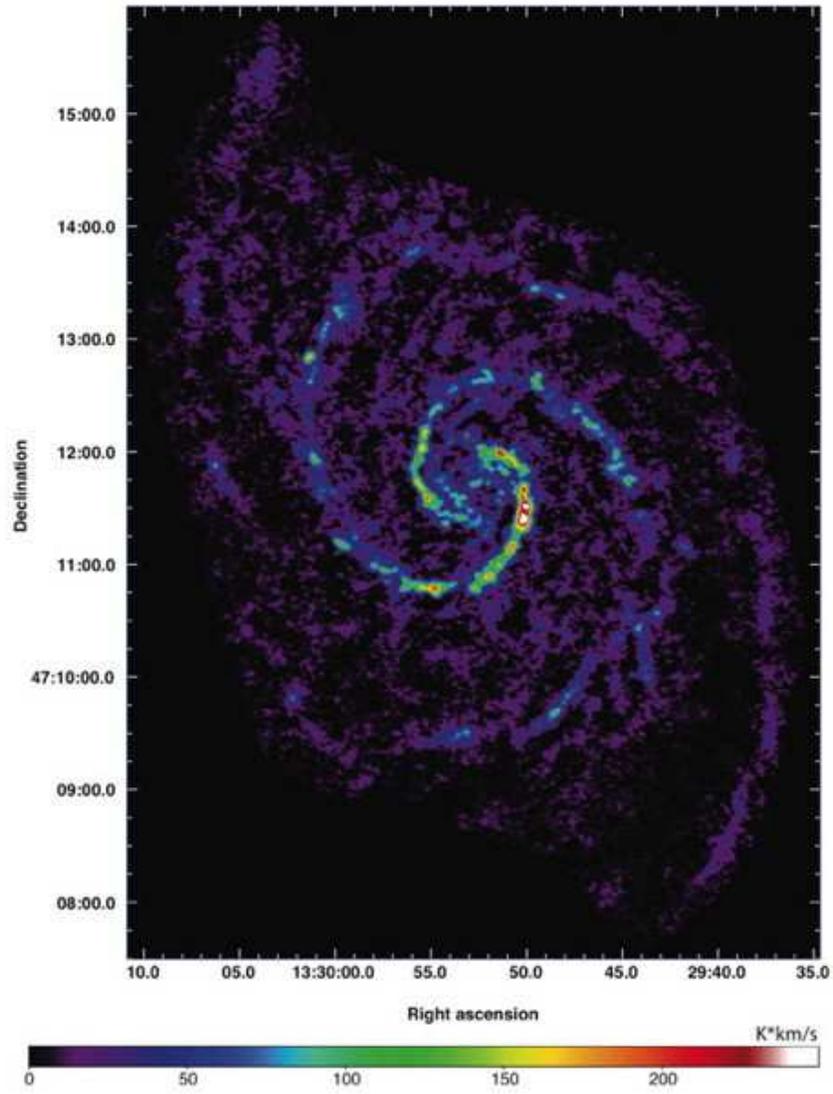}
\caption{CARMA and NRO45 combined CO ($J=1-0$) map of M51 with {\it robust=-2}.
 \label{fig:combmap}}
\end{figure}

\begin{figure}
\epsscale{.80}
\plotone{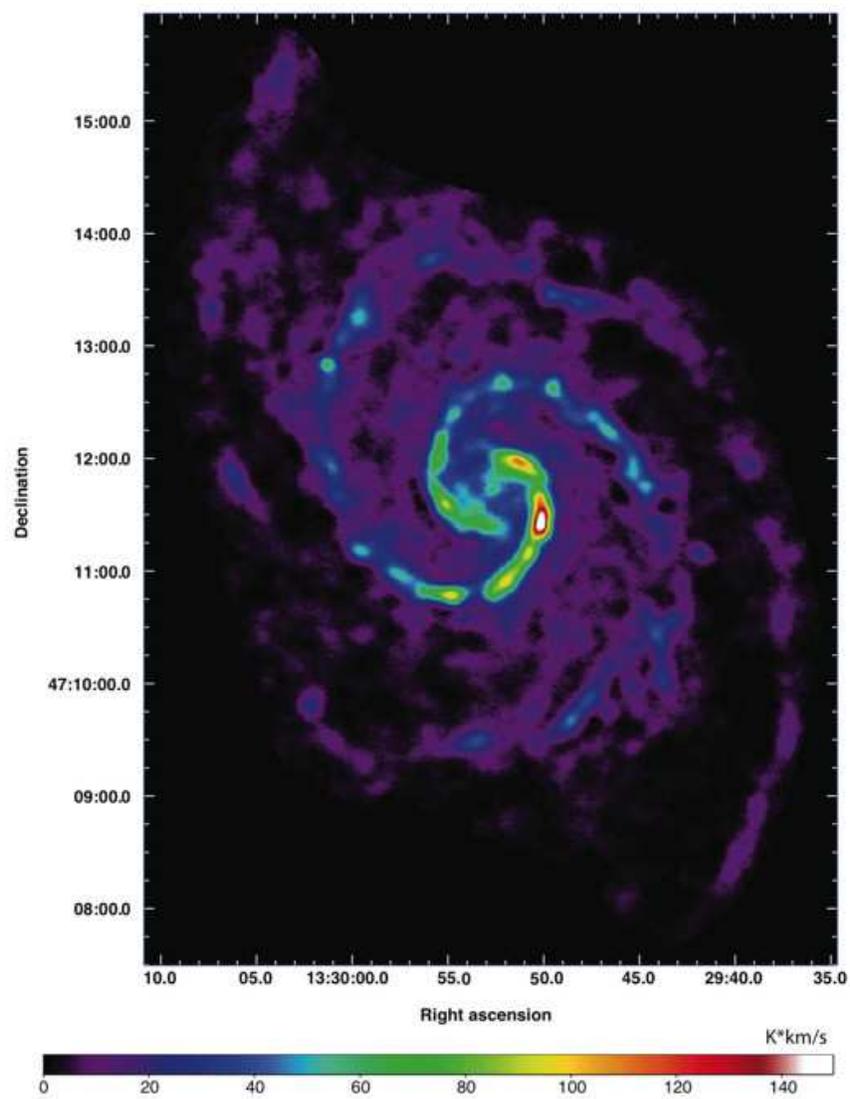}
\caption{The same as Figure 7, but with {\it robust=+2}.
 \label{fig:combmapna}}
\end{figure}

Figure \ref{fig:combmap} and \ref{fig:combmapna} show the CO($J=1-0$) integrated intensity maps of M51,
the combination of CARMA and NRO45 data, with {\it robust} = -2 and +2, respectively.
These maps are made with the "masked moment method" in \citet{adl92}.
We also dropped the low sensitivity region (outer region) of the CARMA mosaic
(see Figure \ref{fig:pointing}). The data with {\it robust} = -2 is used in following discussions,
since it shows finer structures at a higher resolution.

\begin{figure}
\epsscale{1.00}
\plotone{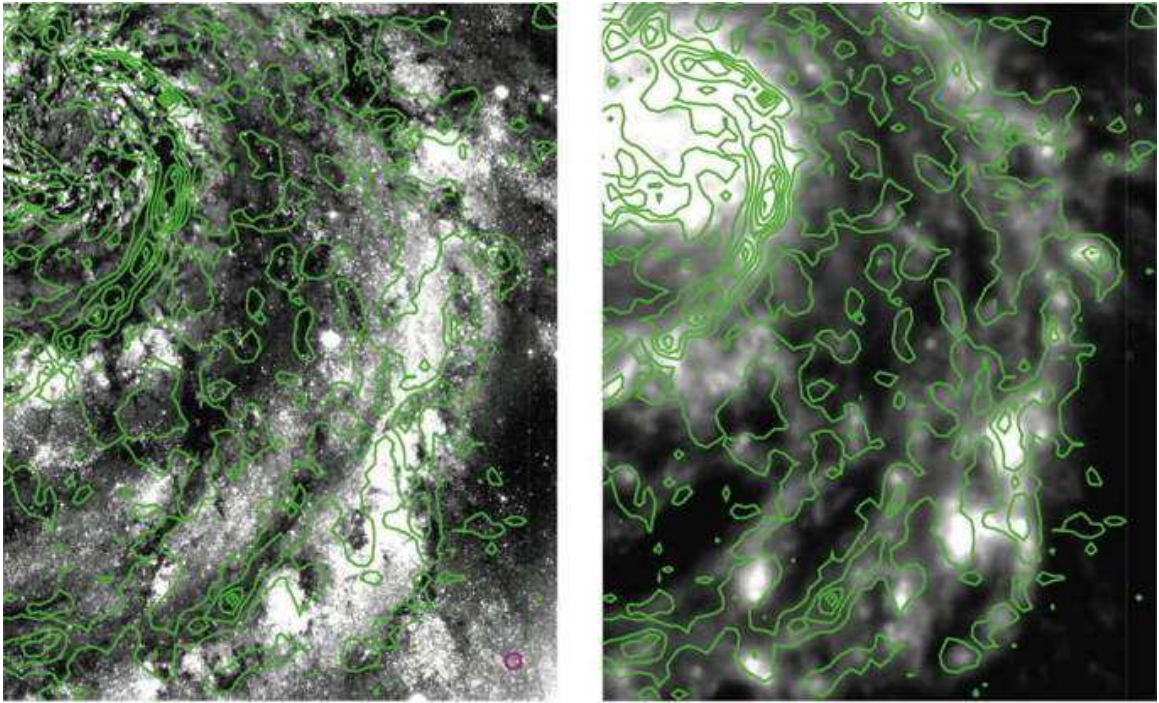}
\caption{CO contours on an {\it HST} $B$-band image ({\it left}) and on a {\it Spitzer} $8\mu m$ image ({\it right}).
The CO image is first smoothed to a $30\arcsec$ resolution to mask out $<3\sigma$ pixels
in the smoothed image. The CO contours are made with the masked {\it robust=-2} map (not with
the integrated intensity map), and are the superposition of the 1.5, 5.0, 10.0, 15.0 $\sigma$ contours
in all $10\kmps$ channels. The extended emissions at the $1.5\sigma$ level are significant in the smoothed image.
 The {\it HST} image is derived by dividing the original image by an axisymmetric luminosity profile to visualize the dark dust lanes. The CO emission coincides with the dust lanes both on bright spiral arms and in dimmer interarm regions, indicating the high fidelity of the CO data. The {\it Spitzer} $8\mu m$ image traces the dense interstellar medium illuminated by UV photons, and the CO coincides with the $8\mu m$ emission as well. The circle at the lower-right corner of the left panel has the diameter of $4\arcsec$ (roughly the size of beam).
\label{fig:hstspitzerco}}
\end{figure}

The combination of CARMA (15 antennas) and NRO45 enables a full census
of the population of giant molecular clouds (GMCs) over the entire galactic disk.
Molecular gas emission in two spiral arms and interarm regions are prominent
in this map. \citet{kod09} showed the distribution of GMCs both in spiral arms and
interarm regions, and the high molecular gas fraction in both regions.
These two results suggest that stellar feedback is inefficient to destroy GMCs
and molecules, which is supported by a recent analysis by \citet{sch10}.
Molecular structures in the interarm regions were often an issue of debate
in previous observations due to poor image fidelity \citep{ran90, aal99, hel03}.
Figure \ref{fig:hstspitzerco} compares the CO distribution with a $B$-band image
from the {\it Hubble Space Telescope} (HST) and an $8\mu m$ image from 
the {\it Spitzer Space Telescope}. Dust lanes in the $B$-band image indicate
the distribution of the dense interstellar medium (ISM), and the $8\mu m$ image
shows the distribution of the PAH (large molecules) illuminated by UV photons
from surrounding young stars. The CO emission coincides very well with
the dust lanes and $8\mu m$ emission in both spiral arms and interarm regions,
which evidences the high image-fidelity over a wide range of flux.

\placefigure{fig:combmap}

\begin{figure}
\epsscale{1.00}
\plotone{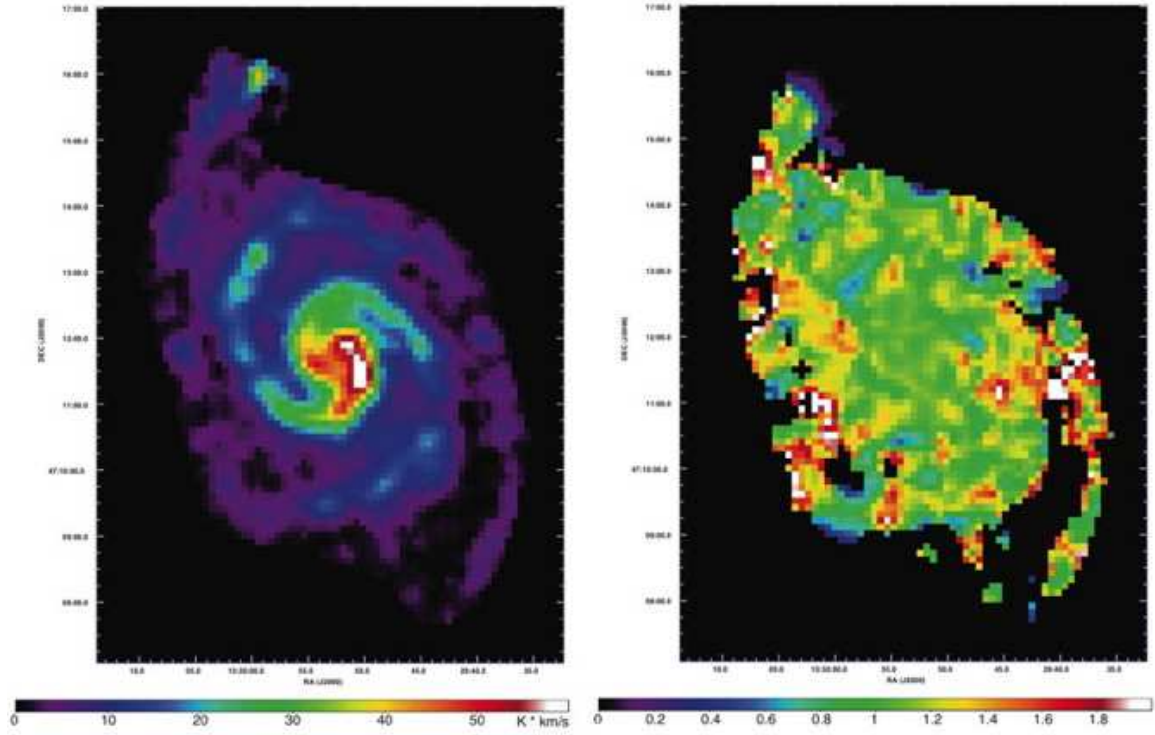}
\caption{{\it Left:} NRO45 CO ($J=1-0$) map.
{\it Right:} Recovered flux map, i.e., the ratio of the combined map over NRO45 map. 
The combined map ($robust=-2$) is smoothed to $\sim 20\arcsec$ resolution for the comparison.
The ratio is $\sim 1$ over the entire map, and the flux recovery is very good.
The companion galaxy NGC 5195 is not in the CARMA velocity coverage.
Note that the error in the ratio map varies across the map, depending on the brightness of emission.
 \label{fig:nrorec}}
\end{figure}

Figure \ref{fig:nrorec} shows the NRO45 map (left)
and the ratio of the combined map (smoothed to $\sim 20\arcsec$ resolution)
over the NRO45 map, i.e. recovered flux map (right).
The recovered flux map shows an almost constant ratio $\sim 1$
over the entire map, and no correlation with galactic structures (i.e., no size
dependence -- in contrast to the dependence expected in pure-interferometer maps).
Some extended CO emission is not significantly detected at the high resolution
of the combined image, but becomes apparent when the image is smoothed.
Note that the companion galaxy NGC 5195 is
not included in the CARMA velocity coverage, nor in the combined map,
though it is in the NRO45 map.

\placefigure{fig:hstspitzerco}
\placefigure{fig:nrorec}


Both the main galaxy NGC 5194 and companion galaxy NGC 5195,
are observed with NRO45.
The total flux of NGC 5194 is $(1.022\pm 0.002 ) \times10^4 \,\rm Jy \cdot km/s$ in the NRO45 map,
which is consistent with the measurement of \citet{hel03}.
With the Galactic CO-to-H$_2$ conversion factor
$X_{\rm CO}= 1.8\times 10^{20} \rm \, cm^{-2} [K \cdot km/s]^{-1}$
and the distance of 8.2 Mpc,
the total molecular gas mass in NGC 5194 is $4.9\times 10^9 \Msun$.
The $X_{\rm CO}$ similar to the Galactic one is found in M51 recently \citep{sch10}.
The total flux of the combined cube is also $1.0\times 10^4 \rm Jy \cdot km/s$,
consistent with the NRO45-only measurement.
The total flux and mass of NGC 5195 is $162 \pm 4 \,\rm Jy \cdot km/s$
and $7.8\times 10^7 \Msun$, respectively. The errors are based on the RMS
from the map, and do not include the systematic error due to the flux calibration
in the CARMA observations ($\sim 15$\%).


\section{Requirements} \label{sec:req}

There are requirements for sampling, field of view, $uv$-coverage, and
sensitivity for single-dish data to be combined
with interferometer data in an optimal manner.

\begin{figure}
\epsscale{.50}
\plotone{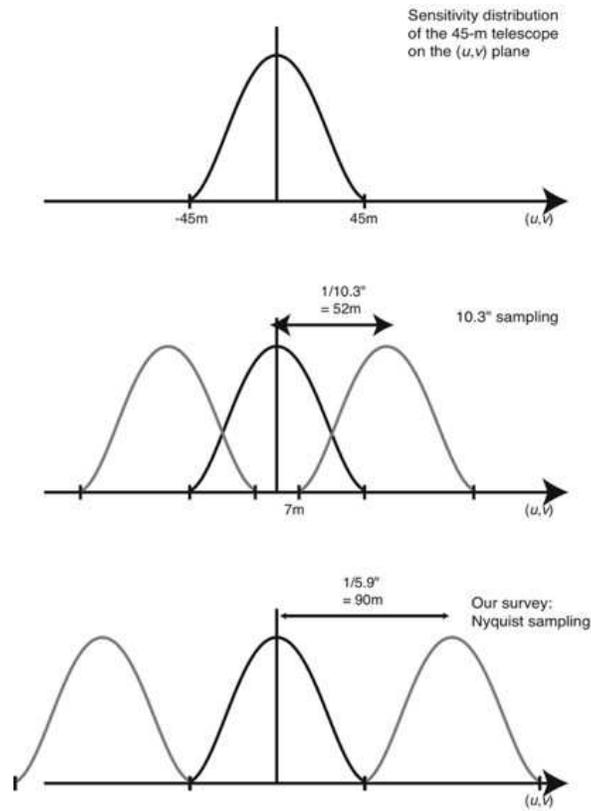}
\caption{Schematic illustrations of the aliasing effect in the $uv$ coverage of the NRO45 map. The alias of the NRO45 data in $uv$ space destroys not only long, but also short baselines. In the case of $10.3\arcsec$ sampling, the baselines as short as 7 m are affected by the alias.
 \label{fig:uvnro45}}
\end{figure}

First, a spatial fine sampling is necessary \citep{vog84}. The half-beam sampling,
a typical practice in most single-dish mapping observations, is not sufficient,
since the aliasing effect destroys visibilities {\it both} at long and very short baselines.
Figure \ref{fig:uvnro45} illustrates the effect schematically: if the spatial sampling
is $10.3\arcsec$ \citep[$=\lambda_{\rm CO} / 52\rm m$, a typical sampling in NRO45
observations; e.g. ][]{kun07}, the tail of the $uv$ distribution leaks into
baselines as short as $\sim 7$ m.
Hence, the Nyquist sample of $11.9\arcsec$ (=$\lambda_{\rm CO}$/45m)
is necessary to properly reproduce visibilities up to the 45 m baseline.
The observing grid and pixel size in the NRO45 map must be at most
$5.96\arcsec$ (Figure \ref{fig:uvnro45}).

The single-dish map should cover an area larger than the area of the
joint map. The deconvolution with the single-dish beam (\S \ref{sec:deconvnro45})
causes artifacts at the edges
of the images. It is ideal to have extra-margins with the width of a few single-dish beam sizes
at each image edge.

The sensitivity match between single-dish and interferometer data should also be
considered in matching their $uv$ coverages; the maximum effective NRO45 baselines are
limited by the matched sensitivity
in our observations. Only the baselines of about 1/4-1/3 of the 45m diameter take practical effect
in the combination.
It is often discussed that a single-dish telescope needs to be about twice larger than
the shortest baseline used in interferometer observations due to uncertainty in
the single-dish beam shape and errors in pointing \citep[see ][ and references therein]{kur09, cor10b}.
In our case, the maximum effective baseline is shorter than this length.
In practice, interferometer data rarely cover the theoretical minimum baseline
(i.e. dish diameter). The long baselines of single-dish data do not have
a sensitivity comparable to interferometer's  (\S \ref{sec:nrouv}).
To avoid a gap in $uv$ coverage without sensitivity loss, the diameter of
the single-dish needs to be 3-4 times larger, unless the receiver of
the single-dish telescope has a significantly higher sensitivity.
The sensitivity match is discussed in detail in Appendix \ref{sec:senmatch}.


\section{Comparisons with Other Methods}\label{sec:othermethods}

Several methods for the combination of single-dish and interferometer data have
been applied at millimeter wavelength. None of the previous data, however,
have a sufficient overlap between single-dish and interferometer $uv$ coverages
(in the sense discussed in \S \ref{sec:deconvnro45}).
The weighting schemes are artificial, rather than based on the sensitivity (i.e., data quality).
Nevertheless, these methods have some advantages in simplicity,
as well as disadvantages in detail.

\citet{sta99} introduce a combination method in the image domain.
This method is adopted for the BIMA Survey Of Nearby Galaxies (BIMA-SONG)
to combine BIMA interferometer with NRAO12 single-dish data \citep{hel03}.
They set the weights to be inversely proportional to the beam area (i.e., one term
in eq. \ref{eq:senim} and \ref{eq:nro45isen}) and add the dirty maps
and beams of BIMA and NRAO12 linearly (eq. \ref{eq:pb3}) to produce
a joint dirty map and beam. The relative weights are
manually and continuously changed with $uv$ distance.
The joint dirty map is then CLEANed with the joint synthesized beam.
This method starts the combination process from images, rather than visibilities,
and is simple. It should be able to use a more natural weighting scheme
(e.g., sensitivity $uv$-distribution based on the beam shape and
eq. \ref{eq:nro45fsen}; see also Appendix \ref{sec:senmatch})
if software is developed.

\citet{wei01} also combine a single-dish map and CLEANed interferometer map.
They deconvolve the single-dish map with its beam pattern
and convolve the result with an interferometer convolution beam,
so that the beam attenuation becomes the same for both single-dish and interferometer images. 
Then, they Fourier-transform both images and replace
the interferometer data with the single-dish data at the central $uv$-spacing.
CLEAN is performed separately for interferometer data alone, which does
not take advantage of the high image fidelity of the combined map.
Having only one control parameter -- the choice of $uv$ range to be
replaced -- can be advantageous.

Visibilities are generated from a single-dish map by several authors
\citep{vog84,tak03,rod08,kur09}. Our method is in this branch.
\citet{hel03} summarize difficulties to set the weights for this combination scheme,
and conclude that it is too sensitive to the choice of parameters.
\citet{rod08} and \citet{kur09} suggest to set the relative weight to obtain
a cleaner synthesized beam shape, which is advantageous in deconvolution
(e.g., CLEAN, MEM).
More specifically, \citet{rod08} set the single-dish weight density in $uv$ space
equal to that of the interferometer visibilities that surround the single-dish
$uv$ coverage. \citet{kur09} adjusted the relative weight to zero out the
total amplitude of  the sidelobes of a synthesized beam.
Our weighting scheme is more intrinsic to each set of data and is based solely
on their qualities; the single-dish weight is independent of the interferometer data
and is set based on the RMS noise of a single-dish map.
The weight is not a parameter of choice.

In pure interferometer imaging, the synthesized beam shape is historically
controlled by changing the weight density in $uv$ space.
The robust parameter \citep{bri95} is a famous example that converts
the weight smoothly from the {\it natural} to {\it uniform} weightings.
Once the weight is set in our method, the robust weighting works even
for the combined data, exactly as designed for pure interferometer data.

\section{Summary} \label{sec:summary}

We describe the CARMA observations at the early phase of its operation,
and the OTF observations with the multi-beam receiver BEARS at NRO45.
The standard reduction of CARMA and NRO45 data are also discussed and
extended to the combined data set case.

We explain the basics of the imaging technique for heterogeneous array data,
and show that the combination of interferometer and single-dish data is
an extension of the imaging of heterogeneous array data.
We introduce a method of combination of interferometer and single-dish data
in $uv$-space. The single-dish map is converted to visibilities in $uv$-space.
The weights of the single-dish visibilities are determined based on the RMS noise of the map,
which is more natural than any other artificial weighting schemes.
The synthesized beam size is determined to conserve the flux between
the dirty beam and the convolution beam. Comparisons with other methods
are discussed. The advantages and disadvantages of those methods are
summarized. In the appendices, we discuss the matching of single-dish
and interferometer sensitivities for the combination of the data.

The resultant map shows the high image fidelity and reveals, for the first time,
small structures, such as giant molecular clouds, both in bright spiral arms and
in faint inter-arm regions \citep{kod09}. From the new map, we calculate
that the total masses of NGC 5194 and 5195 are $4.9\times 10^9 \Msun$
and $7.8\times 10^7 \Msun$, respectively,
assuming $X_{\rm CO}= 1.8\times 10^{20} \rm \, cm^{-2} [K \cdot km/s]^{-1}$.

The combination method is designed on a platform of available software (i.e., MIRIAD)
and generates a finite number of discrete visibilities from a single-dish map.
Future software should enable data manipulation directly on maps (grids)
both in real and Fourier spaces, instead of in visibilities (Appendix \ref{sec:gridbase}).
The weights can be determined on a grid basis, rather than on a visibility basis.
Even in such cases, the weights should be determined from the RMS noise of
the map which are related to the quality of data.

\acknowledgments

We thank Yasutaka Kurono for insightful comments on an early draft and the anonymous referee for useful comments.
We also thank all the CARMA integration team members and the support staff at the Nobeyama Radio Observatory.
The Nobeyama 45-m telescope is operated by the Nobeyama Radio Observatory, a branch of the National Astronomical Observatory of Japan.
Support for CARMA construction was derived from the Gordon and Betty Moore Foundation, the Kenneth T. and Eileen L. Norris Foundation, the James S. McDonnell Foundation, the Associates of the California Institute of Technology, the University of Chicago, the states of California, Illinois, and Maryland, and the National Science Foundation. Ongoing CARMA development and operations are supported by the National Science Foundation under a cooperative agreement, and by the CARMA partner universities.
This research was partially supported by HST-AR-11261.01.

{\it Facilities:} \facility{CARMA}, \facility{No:45m}


\clearpage

\clearpage

\clearpage

\clearpage

\clearpage

\clearpage

\clearpage

\clearpage

\clearpage

\clearpage

\clearpage
\begin{figure}
\epsscale{.80}
\plotone{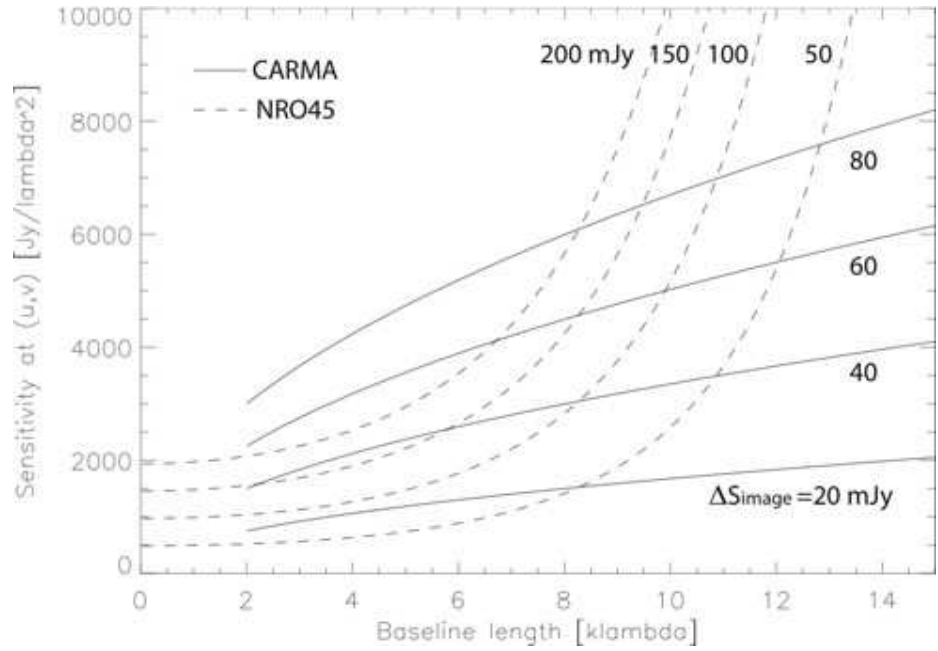}
\caption{Sensitivity distribution with baseline length for CARMA (solid lines) and NRO45 (dashed lines). The labels are the imaging sensitivities of CARMA and NRO45. The minimum and maximum baseline lengths of CARMA are 10 and 300 m, respectively (for C \& D configurations). The sensitivities, pixel sensitivity  $\Delta S^{\rm p}$ ($y$-axis) and imaging sensitivity $\Delta S_{\rm image} (=\Delta S^{\rm i})$, are calculated for the velocity channel width of $\Delta v=10 \kmps$.
 \label{fig:sens}}
\end{figure}


\clearpage
\begin{deluxetable}{lcccccccc}
\tablecaption{Antenna Parameters at 115 GHz\label{tab:ant}}
\tablehead{\colhead{Parameter} & &\colhead{Unit} &  \colhead{Hatcreek (6m)} &  \colhead{OVRO (10m)} &  \colhead{CARMA (6m-10m)} & \colhead{NRO45}  }
\startdata
Main Beam Size      & FWHM                      &  $\rm arcsec$          & 100 & 60 & 77.5$^a$ & 19.7$^b$ \\
Beam Solid Angle   & $\Omega_{\rm b}$ & $\rm arcsec^2$ & $1.51\times 10^4$ & $6.77\times 10^3$  & $1.01\times 10^4$$^a$ & $1.10\times 10^3$$^b$ \\
Quantum Efficiency & $\eta_{\rm q}$         &   \nodata       &   0.87    &  0.87                         & 0.87 & 0.87  \\
Main Beam Efficiency & $\eta_{\rm mb}$  &   \nodata       &  0.41 &  0.61 &  0.50$^a$ & 0.40   \\
Noise Coef. (general)    & $C_{\rm ij}$      & Jy/K & 116.8    &  52.2                         & 78.1$^a$ & 12.0  \\
Noise Coef. (MIRIAD)  & $\rm  JYPERK$ & Jy/K &  145.3    &  65.0                        &  97.2$^a$ & 14.9  \\
\enddata
\tablecomments{Table \ref{tab:ant}}
\tablenotetext{a}{Geometric mean of Hatcreek and OVRO values.}
\tablenotetext{b}{After re-gridding (\S \ref{sec:nro45red}).}
\end{deluxetable}

\clearpage
\appendix

\section{Weight Functions}\label{sec:app}

The dirty image $\bar{I}^{\rm dm}$ and synthesized beam $\bar{B}$ are defined
with a set of visibilities  $V(u,v)$ as

\begin{equation}
\bar{I}^{\rm dm}(l,m) = \int \int V(u,v) W(u,v) e^{2 \pi i (ul+vm)} dudv
\label{eq:dmap}
\end{equation}
and
\begin{equation}
\bar{B}(l,m) = \int \int W(u,v) e^{2 \pi i (ul+vm)} dudv,
\label{eq:bmap}
\end{equation}
where ($u$, $v$) is the coordinates in the $uv$ space.

The sampling and weighting function of visibilities $W(u,v)$ can
be written more explicitly as
\begin{equation}
W(u,v) = \sum^{M}_{k=1} R_k T_k D_k \delta(u-u_k, v-v_k),
\end{equation}
where $T_k$ is the tapering function,
and $D_k$ is the density weighting function \citep[see ][]{tay99}.
$M$ is the number of visibilities obtained in observations.
$T_k$ and $D_k$ are arbitrary functions, and are often used to control
the synthesized beam shape and noise level.
For example, the Gaussian taper is $T_k=\exp(-\sqrt{u_k^2+v_k^2}/2a^2)$
with the half power beam width
$\theta_{\rm HPBW} = \sqrt{2 \ln 2/\pi}/a=0.37/a$ [radian].
The natural and uniform weightings are $D_k=1$ and $D_k=1/N_k$, respectively,
where $N_k$ is the number of visibilities within a pixel in $uv$ space.

$R_k$ is a weight based on noise, and has the relation $R_k=1/\Delta S_k^2$
with $\Delta S_k$ ($=\Delta S^f$ in \S \ref{sec:noise}). The theoretical noise of an image $\sigma$
can be calculated as
\begin{equation}
\sigma = \sqrt{\left( \sum^M_{k=1} T_k^2 D_k^2 R_k\right) \left( \sum^M_{j=1} R_j\right)} / \sum^M_{i=1} T_i D_i R_i.
\end{equation}

\section{Beam Solid Angle}\label{sec:solidangle}

The beam solid angle $\Omega_{\rm A}$ of a synthesized beam (dirty beam) is defined as
\begin{eqnarray}
\Omega_{\rm A} &= & \int \int \bar{B}(l,m) dldm  \label{eq:inte1} \\
	&=& \int \int W(u,v)  \left[ \int \int e^{2 \pi i (ul+vm)} dldm \right] dudv \label{eq:inte2}\\
	&=& W(0,0)
\end{eqnarray}
Eq. (\ref{eq:bmap}) is used bewteen eq. (\ref{eq:inte1}) and (\ref{eq:inte2}).
The bracket in eq. (\ref{eq:inte2}) is a $\delta$-function.
We assumed that the maximum of $\bar{B}(l,m)$ is normalized to 1.

Pure interferometer observations do not have zero-spacing data, and therefore, $\Omega_{\rm A}=0$.
A Gaussian beam $ \bar{B}(l,m) = \exp [-(l^2+m^2)/2 \sigma^2]$ has $W(u,v)=2 \pi \sigma^2 \exp [-(u^2+v^2)/ 2 \sigma_{\rm F}^2]$,
where $\sigma_{\rm F} = 1/ 2 \pi \sigma$. Therefore, $\Omega_{\rm A} = 2 \pi \sigma^2$.

\section{Sensitivity Matching Between CARMA and NRO45}\label{sec:senmatch}
Matching the sensitivities of CARMA and NRO45 is crucial in combination.
The sensitivity requirements of the interferometer and single-dish maps are important for
the observation plan, and a simple way to calculate matching sensitivities is therefore 
important.

One approach is to match the sensitivities in $uv$-space around the $uv$ range (baseline range)
where two data sets overlap. In other words, we want to match the pixel sensitivities $\Delta S^{\rm p}$
of CARMA and NRO45, i.e, their sensitivities at pixel ($u$, $v$).
Among some definitions of sensitivity (e.g., \S \ref{sec:noise}),
the imaging sensitivity $\Delta S^{\rm i}$, i.e., noise fluctuation in the map  (eq. \ref{eq:senim}),
is most used to characterize the quality of map and to estimate the feasibility of
observations. Therefore, we first derive the relation between the imaging and pixel sensitivities
in $uv$ space.

For simplicity, we assume that all CARMA antennas are identical to each other, having exactly
the same $T_{\rm sys}$ and $C_{\rm ij}$. Then, a sensitivity is
\begin{equation}
\Delta S = C_{\rm ij} \frac{T_{\rm sys}}{\sqrt{B \cdot t}}, 
\end{equation}
where $B$ is the channel width and $t$ is the integration time. $\Delta S$ is applied to both CARMA
and NRO45, and can mean one of the following three sensitivities:
fringe sensitivity $\Delta S^{\rm f}$ when $t$ is the integration time of a visibility $t_{\rm vis}$ (eq. \ref{eq:dsk});
imaging sensitivity $\Delta S^{\rm i}$ when $t$ is the total integration time, i.e, $t_{\rm vis} N_{\rm vis}$,
where $N_{\rm vis}$ is the total number of visibilities;
and pixel sensitivity $\Delta S^{\rm p}$ when $t$ is the total integration time of the pixel at ($u$,$v$),
i.e, $t_{\rm vis} n(u,v)$, where $n(u,v)$ is the number of visibilities in the pixel.
Therefore, the imaging and pixel sensitivities are related as
\begin{equation}
\Delta S^{\rm p}(u,v) = \Delta S^{\rm i} \sqrt{\frac{N_{\rm vis}}{n(u,v)}}. \label{eq:senrel}
\end{equation}
Hereafter, we derive the relation between $N_{\rm vis}$ and $n(u,v)$.

The $n(u,v)$ for the interferometer (e.g., CARMA) was discussed by \citet{kur09}. For a target at a reasonably high declination,
synthesis interferometric observations provide a visibility distribution of $n(b) \propto 1/b$, where $b$ is the $uv$ distance
$b=\sqrt{u^2 + v^2}$. Th visibilities (total of $N_{\rm vis}$) are distributed within the minimum and maximum baseline
lengths, $b_{\rm min}$ and $b_{\rm max}$ respectively. From $N_{\rm vis} = \int^{b_{\rm max}}_{b_{\rm min}} n(b) 2\pi b db$,
we derive
\begin{equation}
n(b) = \frac{N_{\rm vis}}{2 \pi (b_{\rm max} - b_{\rm min})} \frac{1}{b}. \label{eq:intnvis}
\end{equation}

The $n(u,v)$ for a single-dish telescope (e.g., NRO45) is determined by the beam shape of the telescope.
We assume a Gaussian beam shape,
$\propto \exp[-(l^2+m^2)/2\sigma^2]$, in sky coordinate ($l$,$m$). The full width half maximum (FWHM) of the beam
is ${\rm FWHM} = 2\sqrt{2\ln 2} \sigma$. The $n(u,v)$ is proportional to the Fourier transformation of the beam shape,
$\propto \exp[-(2 \pi \sigma)^2 b^2/2]$. The total of $N_{\rm vis}$ visibilities are within the $uv$ range from zero
to the antenna diameter $d$. Thus,
\begin{equation}
n(b) = \frac{N_{\rm vis} \cdot 2 \pi \sigma^2}{1-\exp[-(2\pi \sigma)^2 d^2/2]} e^{-\frac{(2\pi \sigma)^2 b^2}{2}}.\label{eq:sinnvis}
\end{equation}

Eq. (\ref{eq:senrel}), (\ref{eq:intnvis}), and (\ref{eq:sinnvis}) give the pixel sensitivity for interferometer $\Delta S^p_{\rm int}(b)$
and single-dish $\Delta S^p_{\rm sd}(b)$. Equalizing these two $\Delta S^p_{\rm int}(b)=\Delta S^p_{\rm sd}(b)$ at $b=b_{\rm overlap}$
where the two $uv$ coverages overlap leads to a relation between the image sensitivities (i.e., RMS map noise) of the interferometer and single-dish.
This relation is a rough measure of the matched sensitivities for the combination of the single-dish and interferometer,
and would be useful in planning observations.
The sensitivity matching can be calculated more accurately with eq. (\ref{eq:senrel}), as performed in \S \ref{sec:deconvnro45},
 if we know an accurate $uv$ coverage $n(u,v)$ of interferometer observations.

Figure \ref{fig:sens} plots the pixel sensitivities $\Delta S^{\rm p}$ of CARMA and NRO45 as function of baseline length $b$
for fixed image sensitivities $\Delta S^{\rm i}$. We set $b_{\rm min}$ and $b_{\rm max}$ to 10 and 300 m
($\sim 4$ and $115 \rm k\lambda$), respectively, for CARMA C \& D-configuratlions.
The CARMA and NRO45 $uv$ coverages overlap significantly between 4 and 10 $\rm k\lambda$.
The NRO45 noise (sensitivity) increases rapidly beyond the baseline length of about half the diameter ($\sim 8 \rm k\lambda$),
and CARMA can complement the $uv$ range beyond that.
The imaging sensitivities of our CARMA and NRO45 observations are
27 and 155 mJy in the velocity width of $10\kmps$, respectively.
Their sensitivities match around $b\sim 4$-$6\rm k\lambda$, within the range where the $uv$ coverages overlap.

\placefigure{fig:sens}

\section{Application To Grid-Based Combination Scheme}\label{sec:gridbase}

The new combination technique discussed in this paper converts a single-dish map
to a finite number of visibilities (discrete data points in $uv$-space).
The weight of each single-dish visibility is determined based on the RMS noise of the
map (i.e., the quality of the data)  using the fringe sensitivity $\Delta S^{\rm f}$ (eq. \ref{eq:nro45fsen}).
In the Fourier transformation, the visibilities are mapped to a grid in $uv$-space,
and the pixel sensitivity $\Delta S^{\rm p}$ is calculated for each pixel of the grid
by summing up the $\Delta S^{\rm f}$ of all the visibilities in the pixel.
The $\Delta S^{\rm p}$ for single-dish data can be calculated directly without going
through visibilities once proper software is developed.

The pixel sensitivity $\Delta S^{\rm p}$ for the single-dish data can be defined with
eqs. (\ref{eq:senrel})(\ref{eq:sinnvis}). The RMS noise of the single-dish map $\Delta S^{\rm i}$
gives the normalization of the equations. Eq. (\ref{eq:sinnvis}) is for a Gaussian beam,
and could be replaced with some other shapes, such as a Fourier transformation of
a single-dish beam or PSF if we have better knowledge of them.
The $\Delta S^{\rm p}$ for the interferometer data should be calculated from the fringe sensitivities
of visibilities $\Delta S^{\rm f}$ using $C_{ij}$ (eqs \ref{eq:dsk}, \ref{eq:cij}) -- mapping the visibilities onto a grid in $uv$-space and
summing up the fringe sensitivities in each pixel.
These $\Delta S^{\rm p}$ naturally set the relative weight of the single-dish and interferometer data.

\end{document}